\documentclass[twocolumn]{aastex62}
\submitjournal{Publications of the Astronomical Society of the Pacific June 2, 2019, Accepted August 19, 2019}
\usepackage{graphicx}
\usepackage{amsmath}
\usepackage{float}
\usepackage{subfigure}
\usepackage{hyperref}

\DeclareMathOperator\erfc{erfc}

\begin{document}

\title{Absolving the \textsc{ssins} of Precision Interferometric Radio Data: A New Technique for Mitigating Faint Radio Frequency Interference}
\author{Michael J. Wilensky}
\affiliation{Department of Physics, University of Washington, Seattle, WA 98195, USA}
\author{Miguel F. Morales}
\affiliation{Department of Physics, University of Washington, Seattle, WA 98195, USA}
\affiliation{Dark Universe Science Center, University of Washington, Seattle, 98195, USA}
\author{Bryna J. Hazelton}
\affiliation{Department of Physics, University of Washington, Seattle, WA 98195, USA}
\affiliation{eScience Institute, University of Washington, Seattle, WA 98195, USA}
\author{Nichole Barry}
\affiliation{School of Physics, The University of Melbourne, Parkville, VIC 3010, Australia}
\affiliation{ARC Centre of Excellence for All Sky Astrophysics in 3 Dimensions (ASTRO 3D), Australia}
\author{Ruby Byrne}
\affiliation{Department of Physics, University of Washington, Seattle, WA 98195, USA}
\author{Sumit Roy}
\affiliation{Department of Electrical and Computer Engineering, University of Washington, Seattle, WA 98195, USA}

\correspondingauthor{Michael J. Wilensky}
\email{mjw768@uw.edu}

\begin{abstract}
        We introduce a new pipeline for analyzing and mitigating radio frequency interference (RFI), which we call Sky-Subtracted Incoherent Noise Spectra (\textsc{ssins}). \textsc{ssins} is designed to identify and remove faint RFI below the single baseline thermal noise by employing a frequency-matched detection algorithm on baseline-averaged amplitudes of time-differenced visibilities. We demonstrate the capabilities of \textsc{ssins} using the Murchison Widefield Array (MWA) in Western Australia. We successfully image aircraft flying over the array via digital television (DTV) reflection detected using \textsc{ssins} and summarize an RFI occupancy survey of MWA Epoch of Reionization data. We describe how to use \textsc{ssins} with new data using a documented, publicly available implementation with comprehensive usage tutorials.
\end{abstract}
\section{Introduction}

Radio frequency interference (RFI) is a ubiquitous problem in radio astronomy. The omnipresence of anthropogenic radio emission and natural emissions from sources such as lightning make it exceedingly difficult to place a serviceable radio telescope in a perfectly radio-quiet location. For instance, the Murchison Widefield Array (MWA)~\citep{Tingay2012, Wayth2018}, which is located in the extremely radio-quiet Murchison Radio Observatory (MRO) in Western Australia, regularly observes ORBCOMM satellite transmissions~\citep{Sokolowski2016}, lightning~\citep{Sokolowski2016}, as well as FM radio signals reflecting off of satellites~\citep{Zhang2018}, meteor trails~\citep{Zhang2018}, and even the moon ~\citep{McKinley2013,McKinley2018}. The MWA also observes digital television (DTV) transmissions~\citep{Offringa2015, Sokolowski2016}, which we show several instances of in this work. 

The brightness of observed RFI signals can vary by orders of magnitude. A survey of the RFI environment of the MRO, including brightness distribution, is presented in ~\citet{Sokolowski2016}, while an MWA RFI occupancy study is presented in ~\citet{Offringa2015}. In ~\citet{Offringa2013}, LOFAR ~\citep{vanHaarlem2013} is used to study RFI brightness for cases where a uniform spatial distribution of RFI emitters is appropriate to assume. Fortunately, a remote telescope such as the MWA does not experience a spatially uniform distribution of RFI emitters\footnote{This can be seen for the MRO by reconciling the observed brightness distribution in ~\citet{Sokolowski2016} with the theoretical brightness distribution of spatially uniform emitters in ~\citet{Offringa2013}.}. However, faint RFI contamination that is below the single baseline thermal noise is still an undeniable reality of such a telescope.

Using the MWA, we study Epoch of Reionization (EoR) cosmology via redshifted 21-cm emission from atomic Hydrogen. Reviews on the subject of 21-cm radiation and its role in cosmology are presented in \citet{Furlanetto2006} and \citet{Morales2010}. The EoR signal is extremely faint relative to the astrophysical foregrounds and RFI signals. Even observed RFI signals that are fainter than the single baseline thermal noise are orders of magnitude brighter than the expected EoR signal. Hence, subthermal RFI excision is a priority in our analysis. 

 In general, RFI mitigation strategies can be deployed at different stages of the interferometric pipeline. Reviews of RFI mitigation schemes are presented in \citet{An2017}, and \citet{Baan2019}. Given the vast data rates of modern radio telescopes, it is common to perform RFI detection offline in the post-correlation stage of the pipeline. Post-correlation mitigation strategies include neural network approaches \citep{Burd2018, Kerrigan2019}, filters in the uv-plane~\citep{Sekhar2018}, variants of fringe filters \citep{Athreya_2009, Offringa2012}, as well as methods that take advantage of the cyclostationarity of common RFI signals \citep{Bretteil2005, HELLBOURG2012}. Several post-correlation detection methods are compared in \citet{Offringa2010}, including surface fitting, singular value decomposition, and combinatorial thresholding algorithms such as \textsc{SumThreshold}. Successful implementations of the \textsc{SumThreshold} method are presented in \citet{Offringa2015} and \citet{PECK2013} using the \textsc{aoflagger}\footnote{\url{https://sourceforge.net/p/aoflagger/wiki/Home/}} and SERPent\footnote{\url{https://github.com/daniellefenech/SERPent}} software packages, respectively. We use an RFI detection algorithm in this work similar to \textsc{SumThreshold}, with slight differences.

Currently, all MWA EoR observations are flagged for RFI during pre-processing using \textsc{aoflagger}. Like many of the state-of-the-art methods referred to above, \textsc{aoflagger} flags a single baseline at a time. Clearly, single baseline flaggers are fundamentally limited to the sensitivity attainable by a single baseline. As we will show, there exists RFI observed by the MWA  that evades detection by \textsc{aoflagger} because it is fainter than the single baseline thermal noise. We offer Sky-Subtracted Incoherent Noise Spectra (\textsc{ssins}) as a flexible, sensitive, and statistically simple RFI analysis pipeline to detect, classify, and excise such faint RFI. \textsc{ssins} was developed for the MWA, and it has also undergone limited testing on the Hydrogen Epoch of Reionization Array (HERA) ~\citep{Deboer2017} and the Long Wavelength Array (LWA) ~\citep{Taylor2012}. 

After explaining the \textsc{ssins} method in detail (\S\ref{sec:method}), we show several common cases of RFI that are successfully flagged by \textsc{ssins} in the MWA EoR highband, which extends from 167.1 Mhz to 197.7 Mhz (\S\ref{sec:comp}). We also image RFI and hypothesize about the source (\S\ref{sec:images}). We then discuss the implementation of \textsc{ssins}, explain how it can be customized for different RFI environments, and summarize RFI mitigation efforts for data used in an EoR power spectrum limit in \citet{Barry2019b} (\S\ref{sec:custom}). Finally, we discuss possibilities for future work (\S\ref{sec:conc}).

\section{The Method}
\label{sec:method}

The measurements from an interferometer can be broken down into three components: the astrophysical sky signals, thermal noise, and RFI. We will first separate the slowly varying sky signal from the modulated RFI (\S\ref{sub:skysubtraction}), then increase our sensitivity to the faint RFI signal through successive integrations and frequency-matched flaggers (\S\ref{sub:incoherentavg}--\ref{sec:Match-Shape}). In  Figures~\ref{fig:hists}--\ref{fig:MS_SSINS} we compare an observation free of RFI (left) with an observation contaminated by faint RFI (right).

\subsection{Sky Subtraction}
\label{sub:skysubtraction}

The first step, sky subtraction, relies on the time variation of the sky signal being slow relative to the visibility cadence. We denote the visibility belonging to the \(ij\) antenna pair, time integration, \(t_n\), frequency, \(\nu\), and polarization, p, as \(V_{ij}(t_n, \nu, p)\). By subtracting data from subsequent time integrations, the majority of the astrophysical sky signal will be removed while leaving much of the modulated RFI and thermal noise. We write the sky-subtracted visibilities as 
\begin{equation}
    \Delta V_{ij}(t_n, \nu, p) = V_{ij}(t_{n+1}, \nu, p) - V_{ij}(t_n, \nu, p).
    \label{eq:vis_diff}
\end{equation}
What remains has a noise-like component and potentially an RFI-like component. 

\begin{figure*}[t!]
    \centering
        \begin{subfigure}[]
        \centering
        \includegraphics[width=0.49\linewidth]{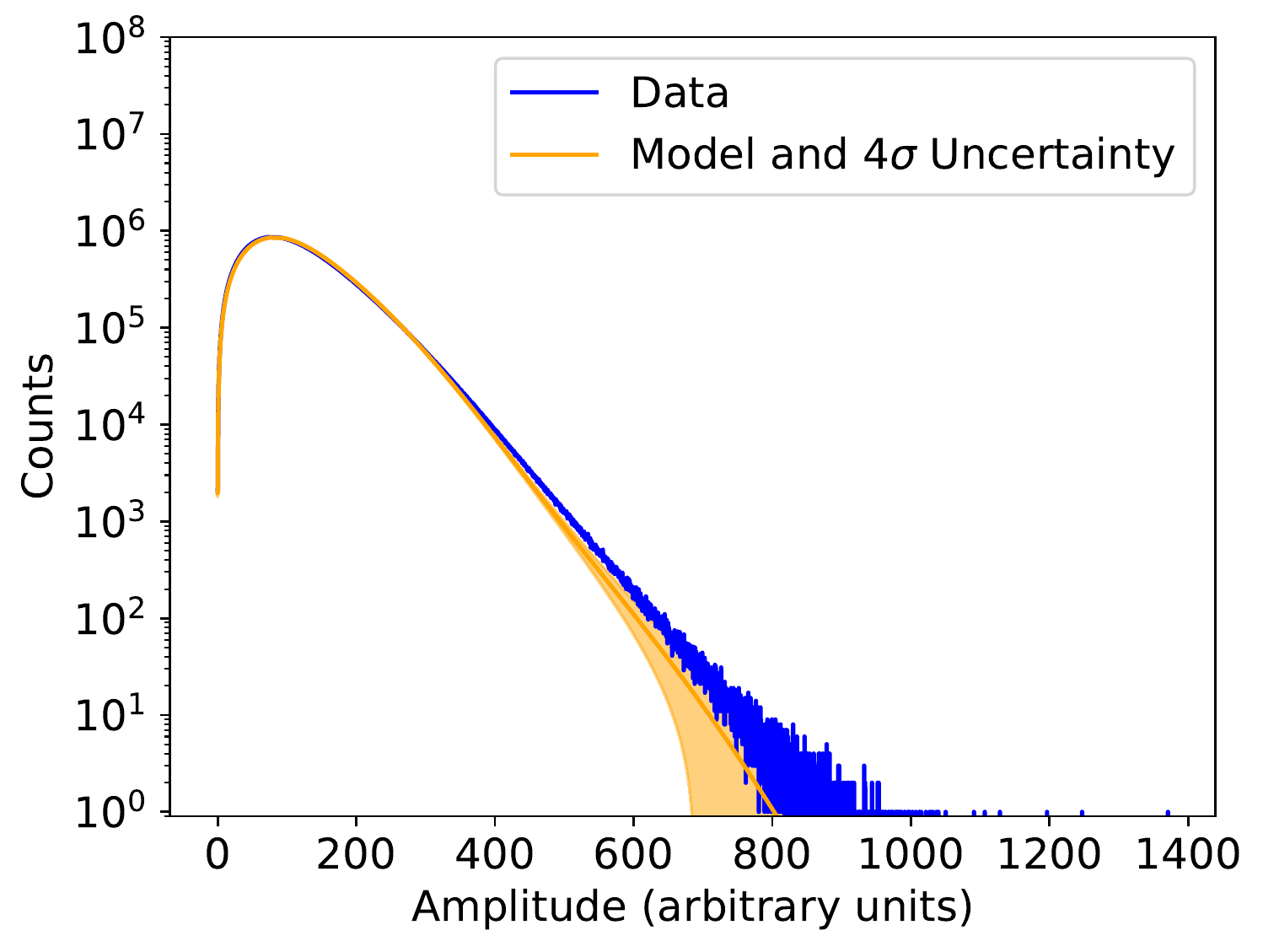}
        \end{subfigure}
        \begin{subfigure}[]
        \centering
        \includegraphics[width=0.49\linewidth]{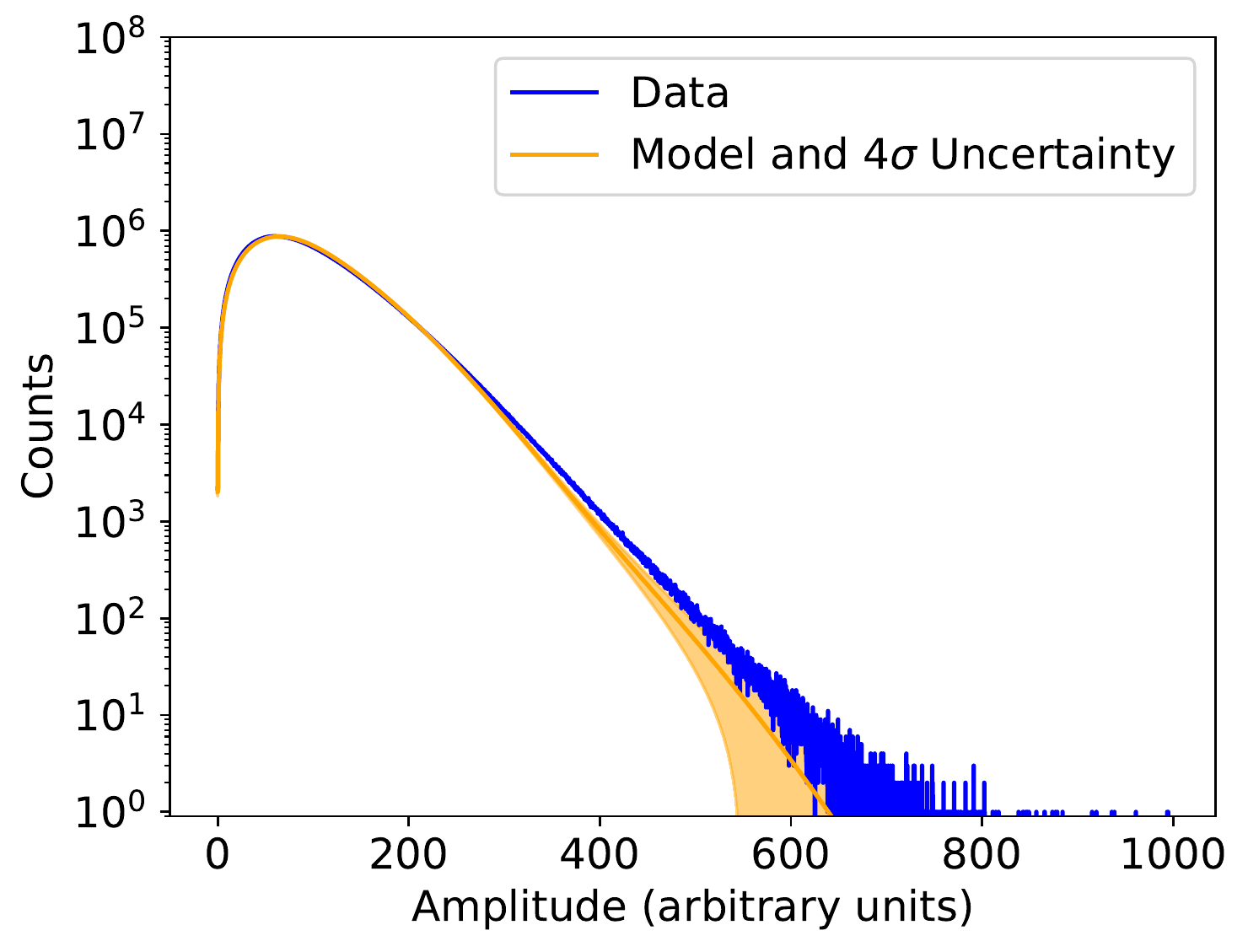}
        \end{subfigure}
    \caption{Histograms of the amplitudes of time-differenced visibilities for two different two-minute MWA observations, similar to Figure 8.2 of~\citet{Barry2018}. (a) belongs to an observation deemed clean by the methods shown in this paper, while (b) belongs to an observation that is shown to have some digital television contamination (\S\ref{sec:Match-Shape}). The measurements are shown in blue, while an accompanying Rayleigh-mixture fit is shown in orange with 4\(\sigma\) error bars. These figures are nearly indistinguishable, despite differences in contamination.}
    \label{fig:hists}
\end{figure*}

To separate the thermal noise from the RFI it is helpful to understand the statistical properties of the thermal noise. For visibility data, the noise is a circular complex Gaussian process. In other words, for each visibility, the real and imaginary components of the noise are independently and identically distributed Gaussian random variables with mean equal to zero. Following standard derivations such as in \citet{Zwillinger2000}, the amplitudes, \(X\), of a circular Gaussian process are Rayleigh distributed: 
\begin{equation}
    f_X(x; \sigma) = \frac{x}{\sigma^2}e^{-x^2/2\sigma^2},
\end{equation}
where \(\sigma^2\) is the variance of the Gaussian which describes the real (or imaginary) component of the process. The mean of the Rayleigh distribution is
\begin{equation}
    E[X] = \sqrt{\frac{\pi}{2}}\sigma,
    \label{eq:rayl_mean}
\end{equation}

\noindent and the variance is 

\begin{equation}
    E[X^2] - E[X]^2 = \frac{4-\pi}{2}\sigma^2.
    \label{eq:rayl_var}
\end{equation}

Figure \ref{fig:hists} shows the Rayleigh-like amplitude distributions of the sky-subtracted visibilities for our two example observations along with a Rayleigh-mixture fit. Because the observed noise has a frequency dependence, each frequency channel was fit independently by maximum likelihood estimation to form the final model fit and 4\(\sigma\) errors.
The similarity of Figures \ref{fig:hists}(a) and (b) shows that the sensitivity of a single visibility is not sufficient to mitigate the faint RFI that we are interested in.

\subsection{The Incoherent Average}
\label{sub:incoherentavg}

We boost our RFI sensitivity by averaging the amplitudes of the sky-subtracted visibilities over all of the baselines in the array, leaving a single dynamic spectrum per polarization. This is not an entirely novel idea. We refer the reader to the last two paragraphs of \S4.2 in \citet{Offringa2015}. Formally, for an array with \(N_A\) antennas, the incoherent average, \(Y\), is
\begin{equation}
    Y(t_n, \nu, p) = \frac{2}{N_A(N_A-1)}\sum_{i=1}^{N_A}\sum_{j > i}^{N_A}|\Delta V_{ij}(t_n, \nu, p)|.
    \label{eq:inc_avg}
\end{equation}
 Since we discard the phase of the sky-subtracted visibilities in the average, we call this an incoherent average. Note that we do not include autocorrelations in the average. We refer to the remaining spectra as sky-subtracted incoherent noise spectra (SSINS), incoherent noise spectra, or just noise spectra.

\begin{figure*}[t!]
    \centering
    \begin{subfigure}[]{\includegraphics[width=0.49\linewidth]{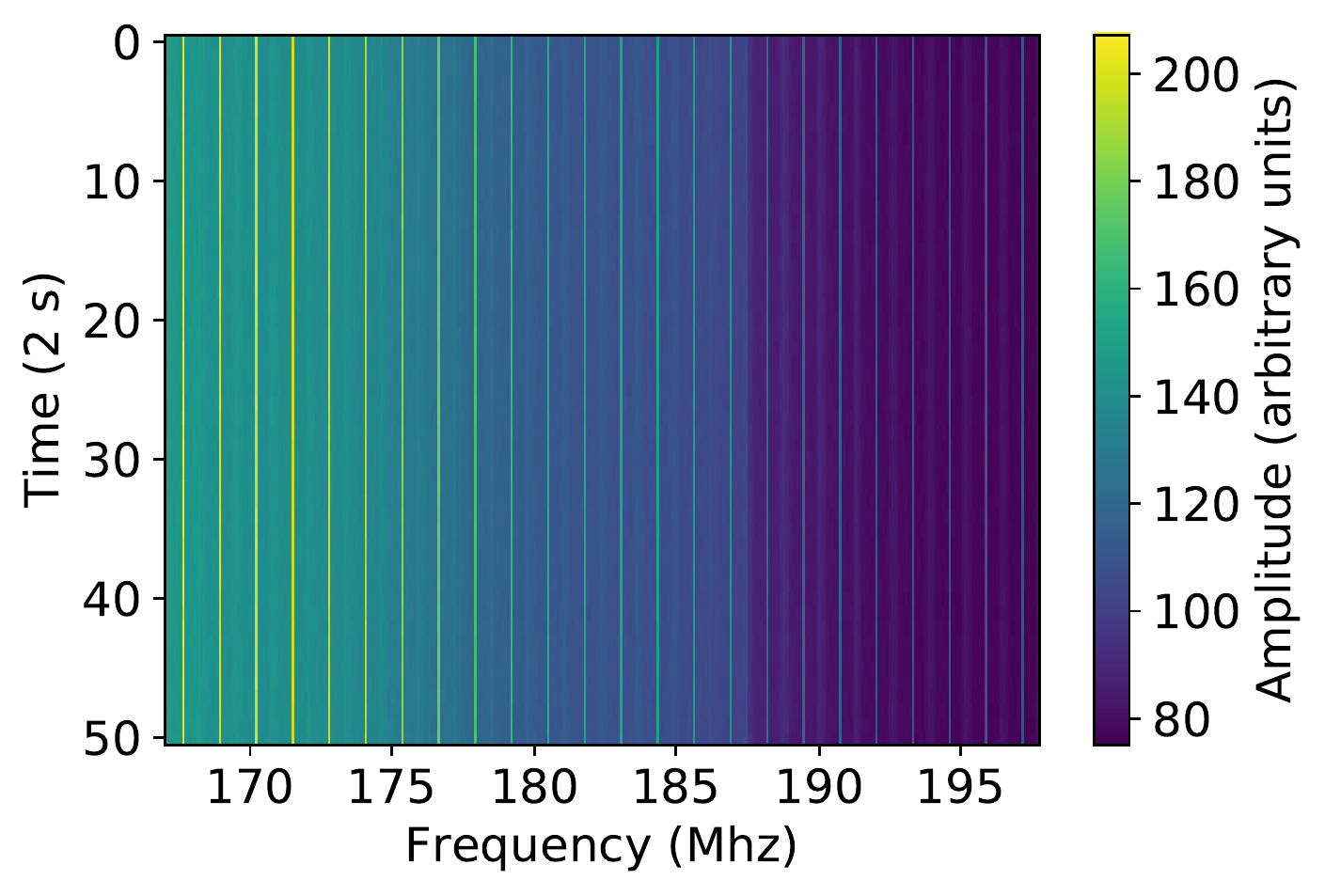}}
    \end{subfigure}
    \begin{subfigure}[]{\includegraphics[width=0.49\linewidth]{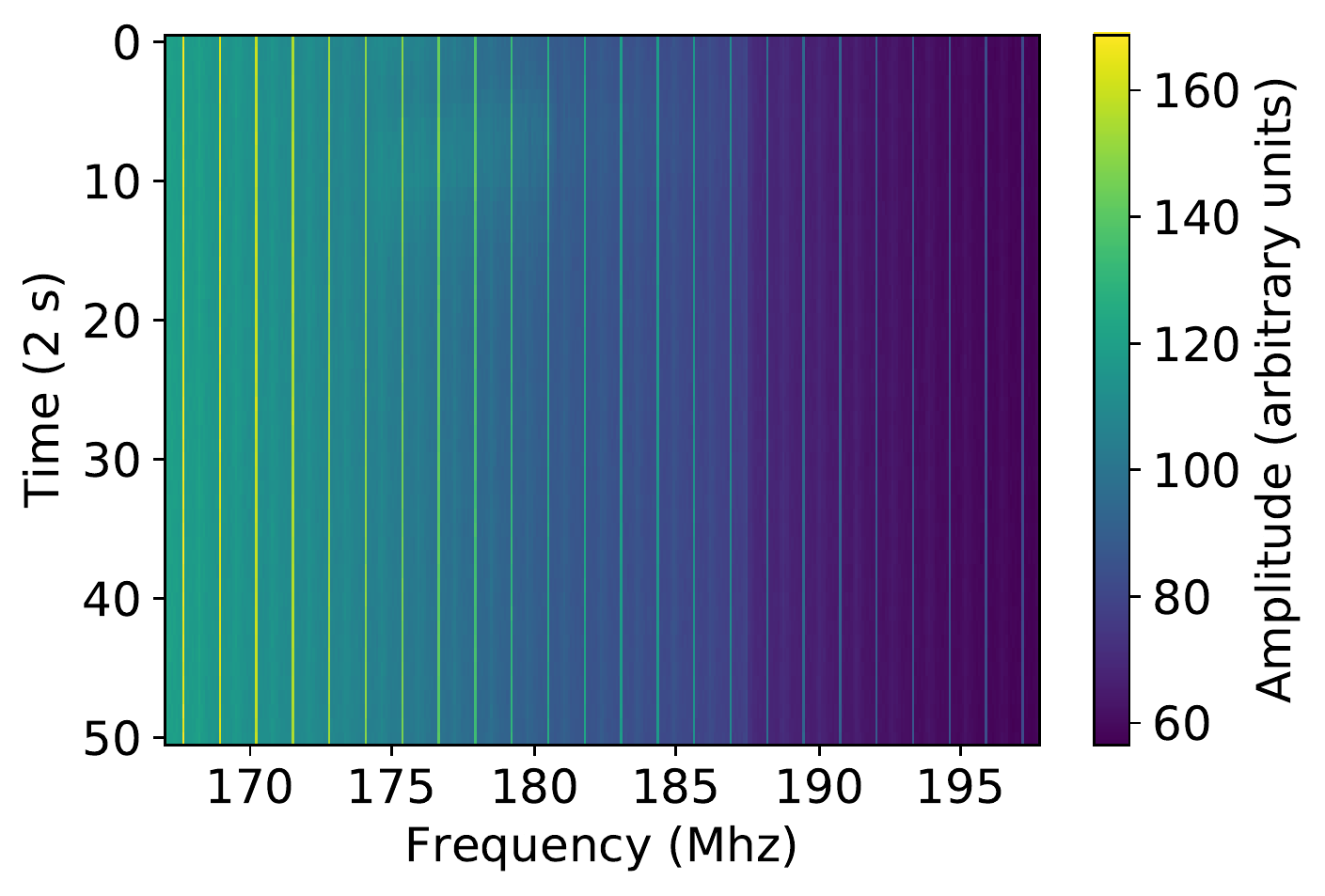}}
    \end{subfigure}
    \caption{The incoherent noise spectrum for the E-W polarization corresponding to the histograms in Figure \ref{fig:hists}, on a two second cadence. The MWA employs a two-stage polyphase filter bank, involving a coarse channelization and then a fine channelization of each coarse channel. The periodic banding in frequency that is seen in these spectra is a result of that filter. The spectrum on the left is clean. However, if one examines the spectrum on the right extremely closely, they may notice a smudge between 174 and 181 Mhz in the first twenty seconds of the observation. This smudge is made much more obvious after the mean-subtraction transformation, shown in Figure \ref{fig:MS_SSINS}, and it is DTV interference.}
    \label{fig:SSINS}
\end{figure*}

Examples of incoherent noise spectra for the same observations as in Figure \ref{fig:hists} are shown in Figure \ref{fig:SSINS}. The spectra in Figure \ref{fig:SSINS}(b) actually show some RFI from digital television broadcasting, although it is extremely difficult to discern from the surrounding noise even in this very sensitive space. In order to determine the nature of such features, we boost the contrast in a way that allows for the rigorous application of statistics.

\subsection{Mean Subtraction}
\label{sub:meansub}

Mean subtraction transforms the data so that the data of a clean observation will be standardized: it will appear as if it were sampled from a zero-mean, unit width Gaussian probability distribution. We describe this process formally using the central limit theorem.

In our context, we take the central limit theorem to say the following (see \citet{Billingsley1995}). Let \((X_1, X_2, ..., X_N)\) be independent and identically distributed random variables with finite mean, \(\mu\), and variance, \(\Sigma^2\). Then, as \(N\) grows large, the sample means given by
\begin{equation}
    S_N = \frac{1}{N}\sum_{k=1}^NX_k
\end{equation}
converge in distribution to a normally distributed random variable of mean, \(\mu\), and variance, \(\Sigma^2/N\). Let us assume the thermal noise to be independent between baselines and ignore baseline-to-baseline noise variation\footnote{Statements of the central limit theorem exist for non-identically distributed sequences, which would be important here if baseline-to-baseline variation in noise levels were a dominant effect. See \citet{Billingsley1995}.}. Having averaged over so many baselines\footnote{Over 8000 for the MWA} in Equation \ref{eq:inc_avg}, the central limit theorem states that the thermal background in the incoherent noise spectrum will be very nearly normally distributed at each frequency with a mean described by Equation \ref{eq:rayl_mean} and variance described by Equation \ref{eq:rayl_var}, divided by the number of baselines in the array, which we denote \(N_{bl}\):
\begin{equation}
    N_{bl} = \frac{N_A(N_A - 1)}{2}.
\end{equation}
Note that the Rayleigh distribution has only a single parameter, so we can actually relate these two quantities. Using Equations \ref{eq:rayl_mean} and \ref{eq:rayl_var}, and writing the underlying mean for each frequency as \(\mu_\nu\), we write the underlying thermal background distribution for each frequency as
\begin{equation}
    f_Y(y; \mu_\nu, N_{bl}) = \sqrt{\frac{N_{bl}}{2\pi C \mu_\nu^2}}\exp\bigg[{-\frac{N_{bl}}{2C\mu_\nu^2}(y-\mu_\nu)^2}\bigg],
\end{equation}
where 
\begin{equation}
    C = \frac{4}{\pi} - 1
\end{equation}
is the ratio of the Rayleigh variance to the square of its mean. 

At this stage, the background distribution for the entire spectrum is a mixture distribution of all the frequencies. If for each frequency we subtract the mean and normalize with respect to the standard deviation, then the background for the entire spectrum will be described by a single distribution (the standard normal distribution), thus simplifying the problem. 

If there are sufficiently many time samples in an observation, then an estimation of the means per frequency can be obtained simply by taking the mean of the spectrum in time:
\begin{equation}
    \hat{\mu}(\nu, p) = \frac{1}{N_t - 1}\sum_{n=1}^{N_{t} - 1}Y(t_n, \nu, p). 
\end{equation}
Writing the mean only as a function of frequency and polarization of course assumes that the thermal process is at least wide-sense (weakly) stationary in time. In the event that there is some drift, then a trend line or trend polynomial can be calculated for each frequency and polarization. For now, we will work with the simplest, most prevalent case.

\begin{figure*}[t!]
    \centering
    \begin{subfigure}[]{\includegraphics[width=0.49\linewidth]{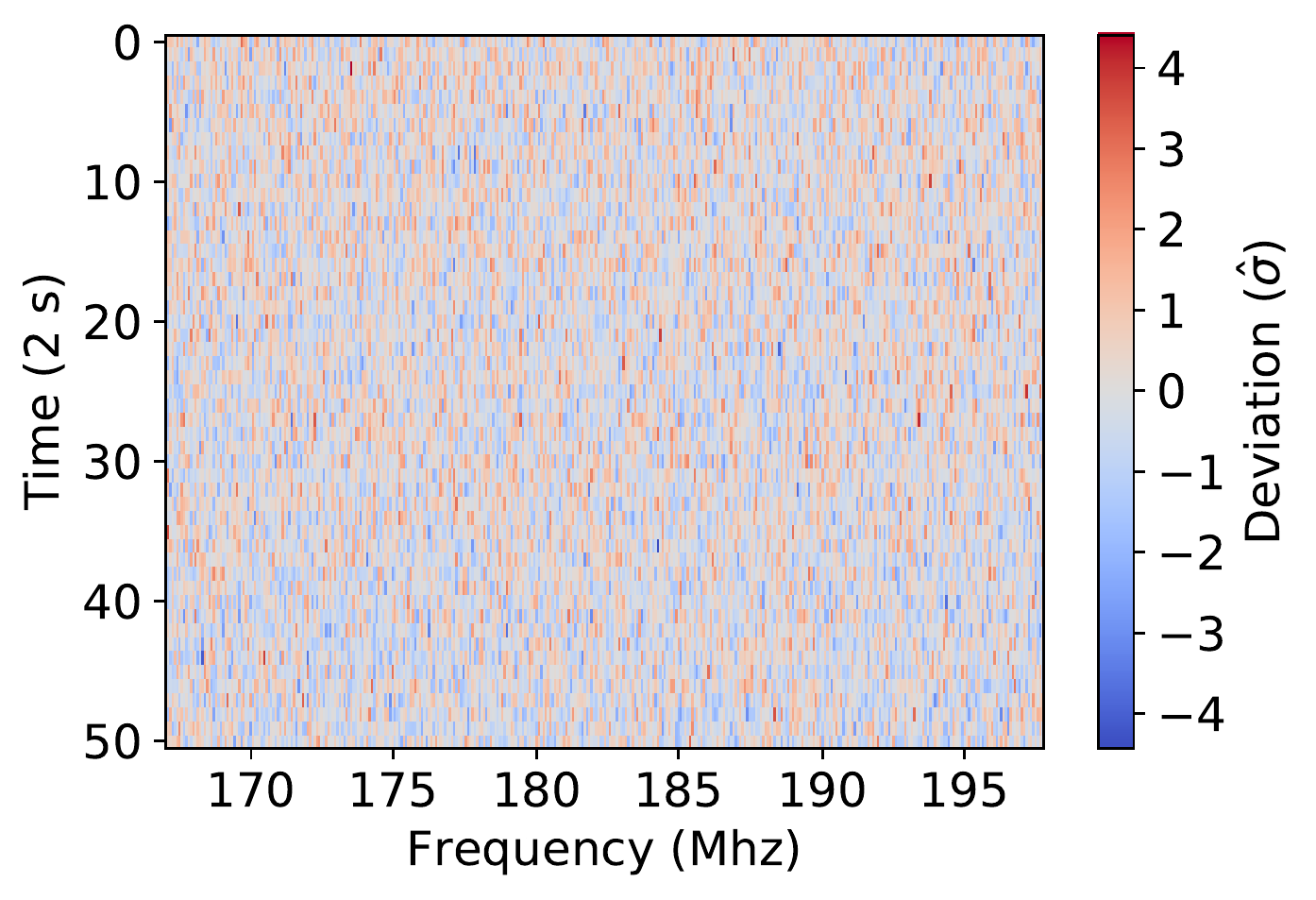}}
    \end{subfigure}
    \begin{subfigure}[]
        {\includegraphics[width=0.49\linewidth]{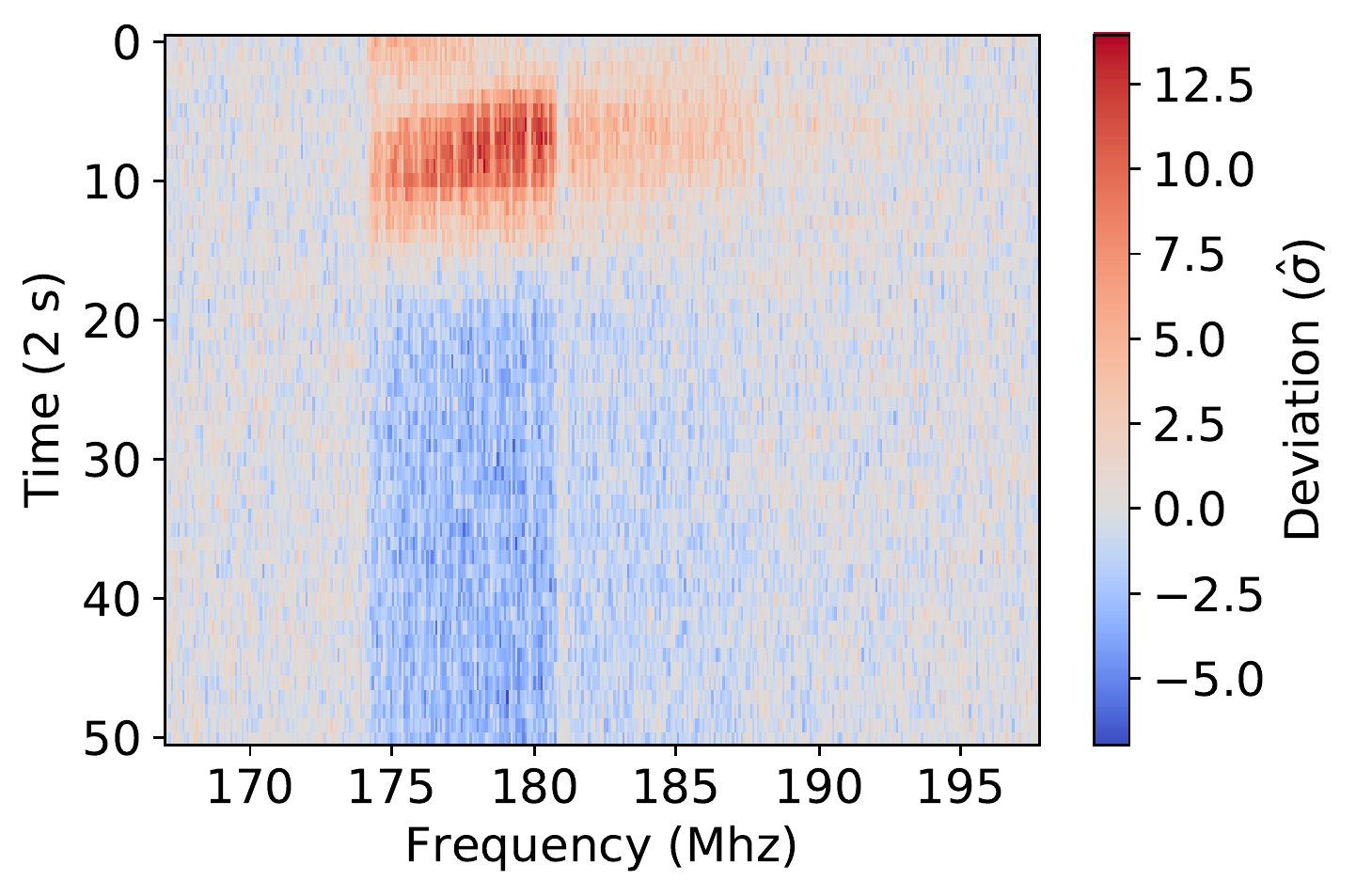}}
    \end{subfigure}
    \centering
    \begin{subfigure}[]{
        \includegraphics[width=0.48\linewidth]{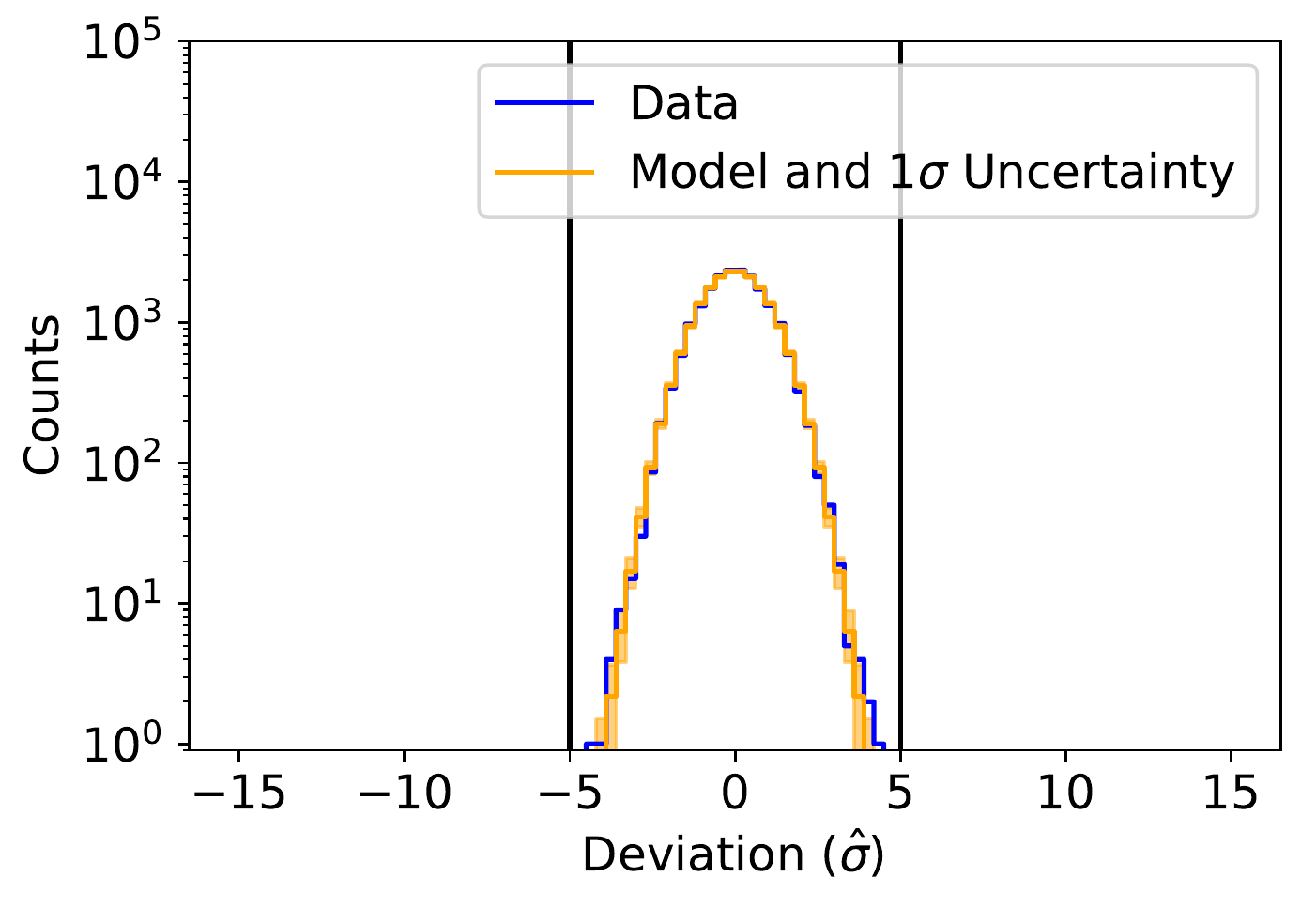}}
    \end{subfigure}
    \begin{subfigure}[]{
        \includegraphics[width=0.48\linewidth]{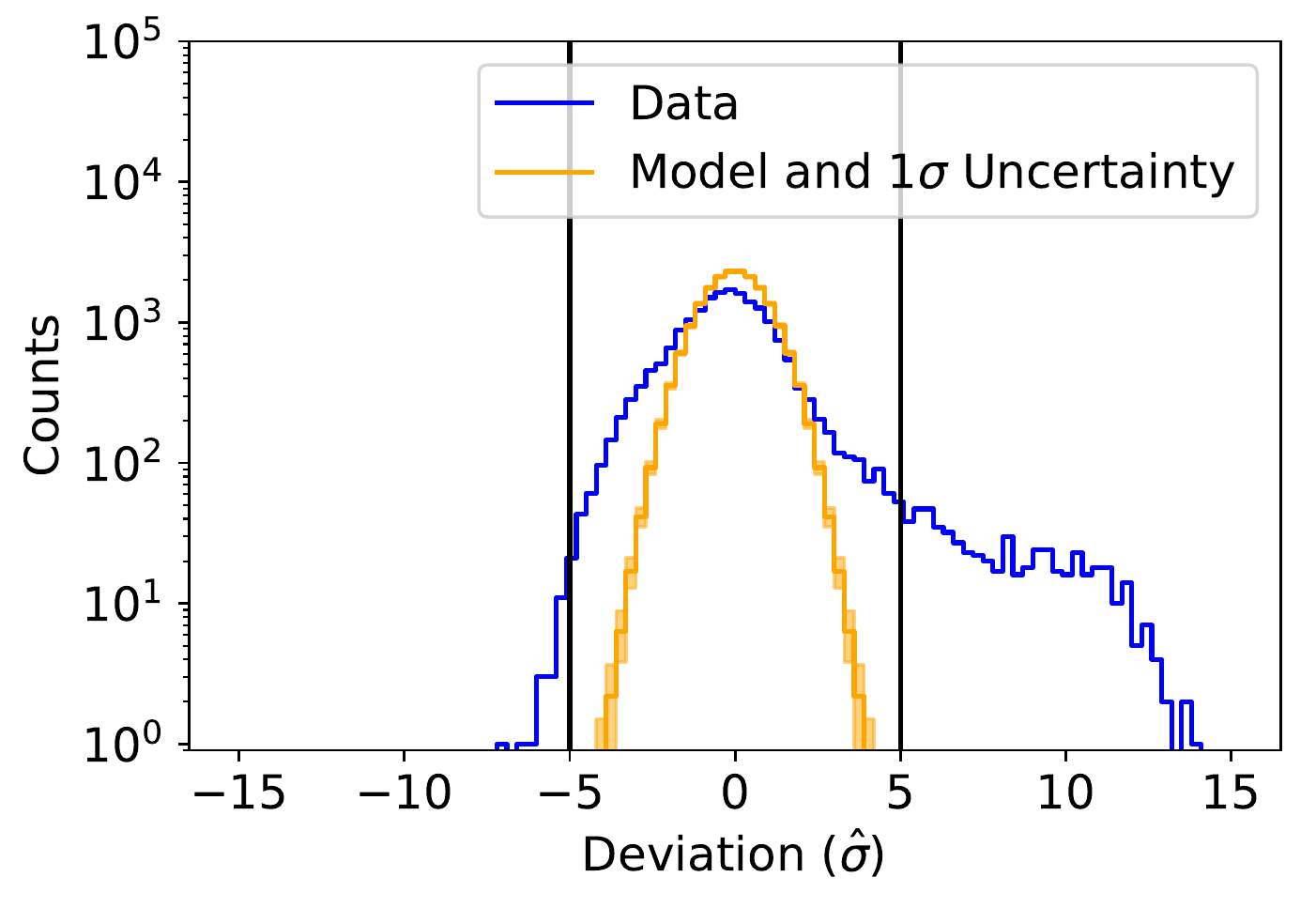}}
    \end{subfigure}
    \caption{Top Row: the mean-subtracted incoherent noise spectra for a single polarization, corresponding to the data in Figure \ref{fig:SSINS}. Bottom Row: Histograms for the mean-subtracted incoherent noise spectra in the top plots, shown with vertical lines that demarcate a point beyond which very few outliers are expected at this data volume. The spectrum in (a) is clean. There are no discernible features such as clustering in time or frequency and the extent of the data is within the range of outliers expected for a data volume of this size. The spectrum in (b), however, features a noticeable cluster of positive outliers that extends from 174-181 Mhz in the beginning of the observation, and a slightly less noticeable one from 181-188 Mhz. Not only is this clustering antithetical to stationary noise, but the data in the brightest feature has outliers as strong as 14\(\hat{\sigma}\), which is not expected at this data volume. In correspondence with this reasoning, the purportedly clean observation in (c) looks highly Gaussian, while the contaminated observation in (d) is clearly deviating from the thermal model.}
    \label{fig:MS_SSINS}
\end{figure*}

Now we consider the quantity
\begin{equation}
    Z(t_n, \nu, p) = \frac{Y - \hat{\mu}}{\sqrt{C\hat{\mu}^2/N_{bl}}}
    \label{eq:mean_sub}
\end{equation}
We call this the mean-subtracted incoherent noise spectrum. Here, for each data point, we have subtracted the estimated mean and normalized with respect to the estimated standard deviation in the corresponding frequency channels and polarizations. Should the observation be totally clean of RFI and if sky-subtraction indeed fully removed the sky from the data, then the remaining data, which will be purely thermal, ought to be very nearly distributed according to
\begin{equation}
    f_Z(z) = \frac{1}{\sqrt{2\pi}}e^{-z^2/2},
\end{equation}
otherwise known as the standard normal distribution. In other words, the quantity in Equation \ref{eq:mean_sub} is a z-score for each sample in the spectrum.

An example of mean-subtraction results following from Figures \ref{fig:hists} and \ref{fig:SSINS} is shown in Figure \ref{fig:MS_SSINS}. The accompanying histograms are plotted along with a standard normal distribution. The clean observation in Figure \ref{fig:MS_SSINS}(c) conforms to the standard normal distribution exceedingly well, while the contaminated observation in Figure \ref{fig:MS_SSINS}(d) has outliers well beyond what is expected, indicating highly non-thermal behavior. Recall that in Figure \ref{fig:hists}, it was nearly impossible to discern any difference between these two observations.

The mean-subtracted spectra are very useful for highlighting RFI that is only marginally brighter than the surrounding thermal noise after incoherently averaging over all baselines. In order to programmatically identify and flag RFI, we deploy a frequency-matched flagger in the mean-subtracted spectrum.

\subsection{The Frequency-Matched Flagger}

\label{sec:Match-Shape}

We extend the probability theory from the previous section to develop a frequency-matched flagger. We introduce this development by showing the results of thresholding on a single-sample basis, and then we show how information from multiple samples can be incorporated.

If the total data volume is \(M\), then the expected number of outliers in a clean mean subtracted spectrum, \(N_{out}\), beyond a certain threshold, \(\tau\), is 
\begin{equation}
    N_{out} = M\erfc\bigg(\frac{\tau}{\sqrt{2}}\bigg),
\end{equation}
where erfc is the complementary error function:
\begin{equation}
    \erfc(x) = \frac{2}{\sqrt{\pi}}\int_x^{\infty}dte^{-t^2}.
\end{equation}
This includes both positive and negative outliers. For a typical two-minute MWA EoR highband observation, the number of expected outliers when \(\tau=5\) is about 0.05. In other words, about one in twenty observations ought to have just a single sample of that strength or greater. Statistically speaking, samples of that strength are exceedingly unlikely to be thermal and so we can be confident that those samples are contaminated. 

It is tempting to flag all data beyond the significance threshold initially, but this can be problematic when there is RFI contamination. Recall that the estimated mean and standard deviation are determined by a time-average in each frequency channel, as described in Equation \ref{eq:mean_sub}. If RFI that is brighter than the thermal noise contaminates some times, then the mean estimate for the contaminated channels will be skewed upward relative to an estimate drawn from only the clean data. As a result, some clean data may appear to be outlying in the negative direction beyond the significance threshold.
For example, consider the bright red cluster followed by the blue trough in Figure \ref{fig:MS_SSINS}(b). As shown in Figure \ref{fig:MS_SSINS}(d), some data in this observation lies beyond the threshold in the negative direction when the z-scores are initially calculated. Flagging everything beyond the threshold initially will inevitably flag some clean data in this case. We can decrease this type of overflagging by taking an iterative approach to flagging, wherein the RFI contamination of the mean estimate is progressively removed in each iteration. This also allows us to probe deeper into the observation for fainter RFI in each iteration. 

\begin{figure}
     \centering
     \includegraphics[width=\columnwidth]{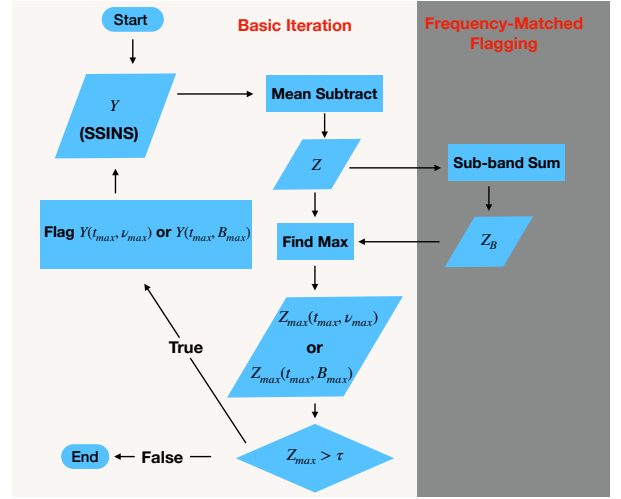}
     \caption{A flowchart of the iterative flagging procedure detailed in \S\ref{sec:Match-Shape}. The white area shows the basic iteration performed on the z-scores calculated in Equation \ref{eq:mean_sub}, while the grey shows two additional steps that are included when frequency-matched flagging is implemented with the z-scores calculated in Equation \ref{eq:Z_B}.}
     \label{fig:flowchart}
 \end{figure}

\begin{figure*}
    \centering
    \begin{subfigure}[]{
        \includegraphics[width=0.48\linewidth]{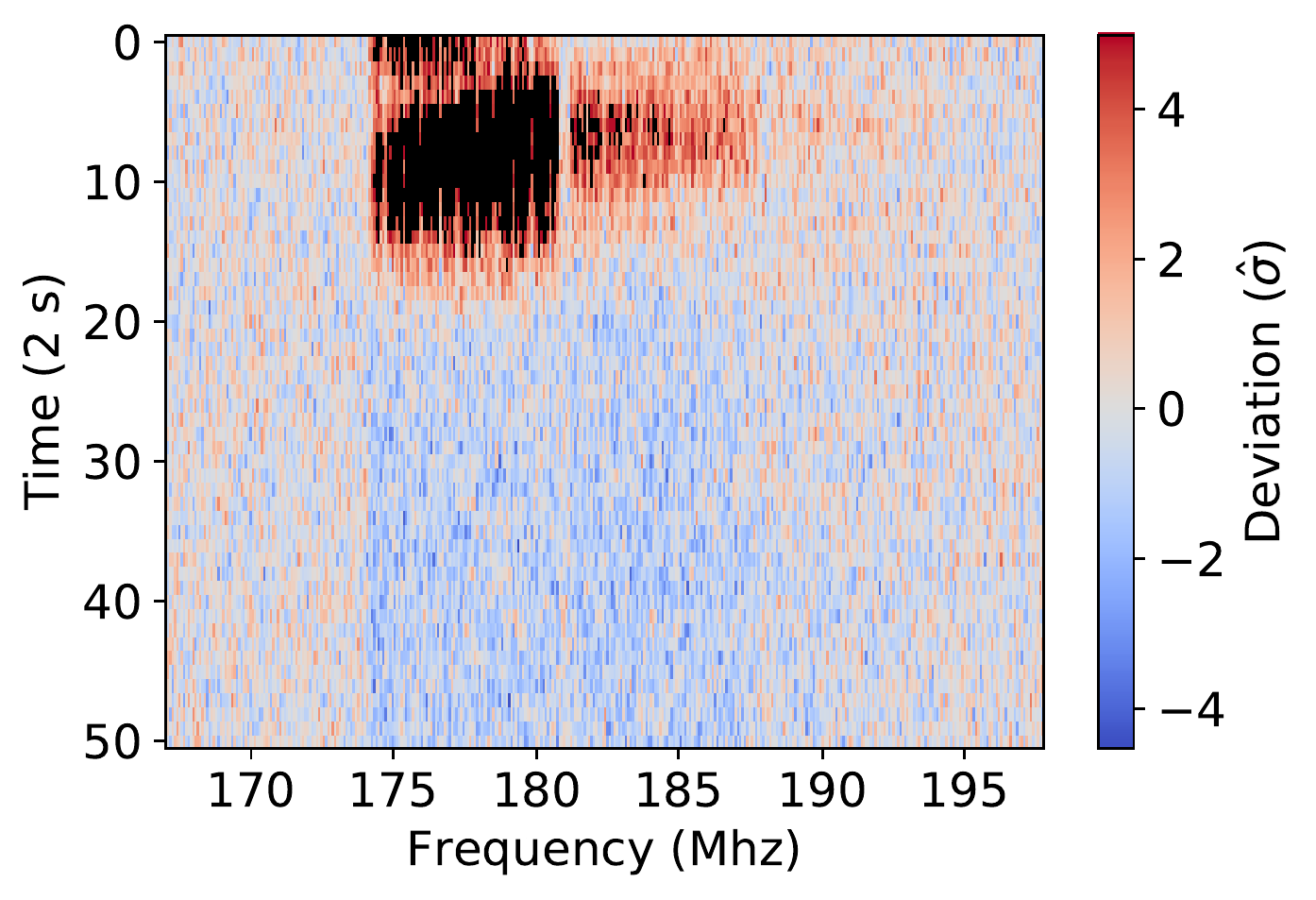}}
    \end{subfigure}
    \begin{subfigure}[]{
        \includegraphics[width=0.48\linewidth]{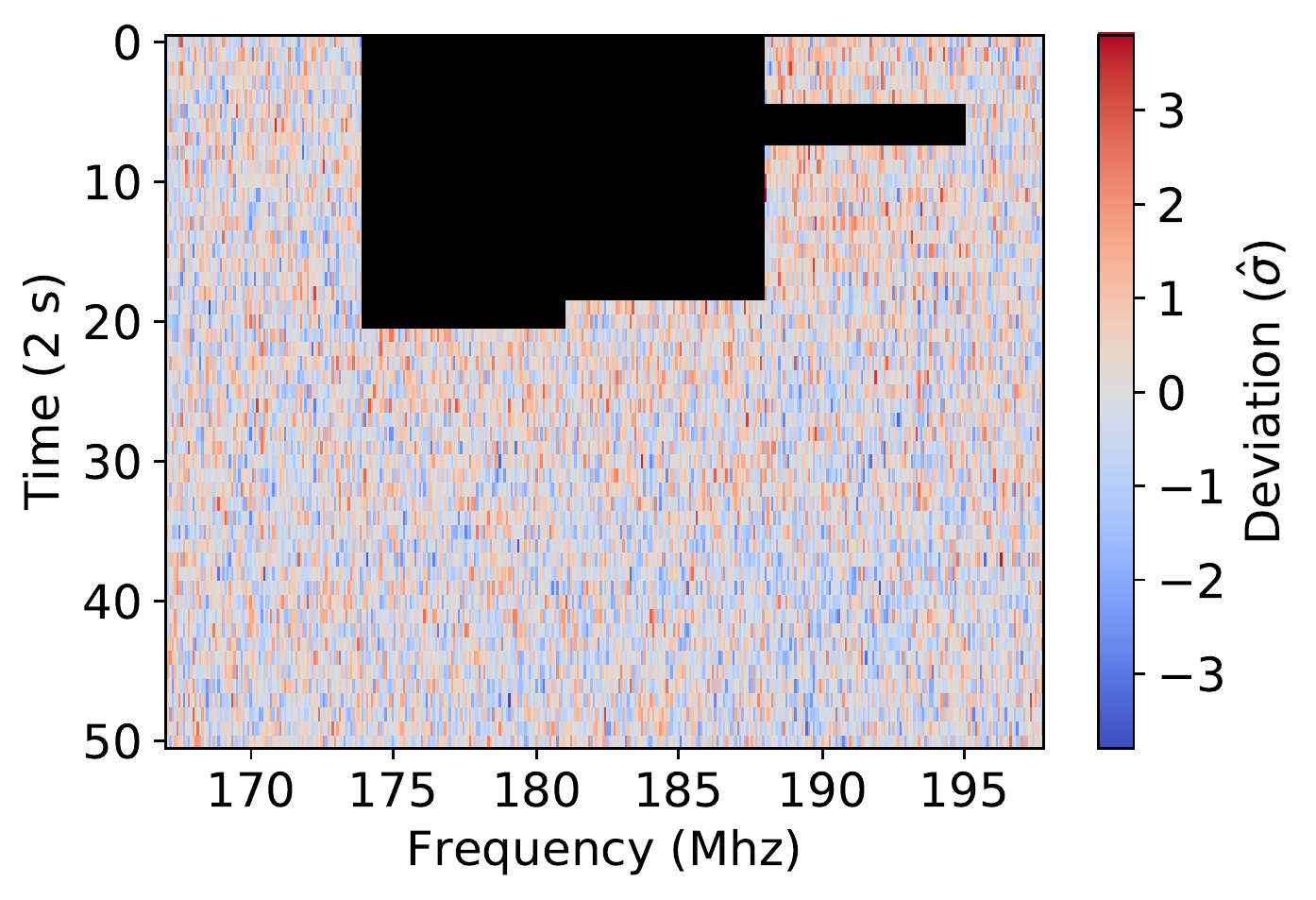}}
    \end{subfigure}
    \caption{A comparison of single-sample thresholding and frequency-matched flagging on the contaminated observation of the previous figures. Black data is flagged. In (a), we show the results of iteratively flagging single-sample outliers beyond the \(\tau=5\) threshold. Leftover RFI persists, since some RFI is beneath this threshold on a single-sample basis. In (b), we show the results of frequency-matched flagging, where we have used bandwidth information about Western Australian DTV to inform the flagger. No remaining interference can be seen in the spectrum. A movie of each flagging iteration is available in the ancillary files.}
    \label{fig:Filt_Demo}
\end{figure*}

 We describe the basic iterative flagging procedure below and present a flow chart in Figure \ref{fig:flowchart}. First, we calculate the z-scores of each noise spectrum sample using mean-subtraction.  Next, we identify the time and frequency of the strongest outlier in the mean-subtracted spectrum that is beyond the threshold in either the positive or negative direction. Then, we flag this time and frequency in all polarizations regardless of the polarization in which the outlier was found. This is done in case that RFI is present in other polarizations at extremely faint levels; it is often the case that RFI is not exactly polarized along the directions measured by the antennas. We then repeat this process until no more outliers exist beyond the threshold. The results of this process using \(\tau=5\) on the contaminated observation shown in previous figures is shown in Figure \ref{fig:Filt_Demo}(a). In this figure, we deviate from the previous convention in Figures \ref{fig:hists}-\ref{fig:MS_SSINS}. Now, the same observation is in both panels. We are comparing differences in flagging between the single-sample outlier iteration detailed above and a frequency-matched flagger, detailed in the main text below. We can see that there appears to be incomplete flagging in the feature noticed in Figure \ref{fig:MS_SSINS}(b), and that another feature immediately adjacent to it of similar width persists in the observation at fainter levels. Any individual remaining sample in these features lies beneath the significance threshold, and so cannot be caught by the method above at the threshold set. To boost our sensitivity to faint features such as these, we combine samples of the mean-subtracted spectrum across frequencies.
 
So far, we have only relied on knowledge of the thermal background without including any specific information about the particular source of the RFI. In the examples shown, data are taken in Western Australia over the frequencies 167-197 Mhz. A subset of the Western Australian digital television channels are broadcast in this observing band and are broadcast 7 Mhz wide, adjacent to one another\footnote{\url{https://www.acma.gov.au/\~/media/Licence-Issue-and-Allocation/Publication/pdf/TVRadio\_Handbook\_Electronic\_edition-pdf.pdf?la=en}}. The broad features shown in the previous figures correspond to DTV channels 6 and 7. The MWA also sometimes observes DTV channel 8. We can hunt for these particular contaminants by summing the mean-subtracted spectrum over the frequencies belonging to a particular type of contaminant. A similar targeted sub-band summing method was employed in \S4 of \citet{Offringa2015} exactly for DTV interference. 

Formally, if an RFI signal occupies a sub-band, \(B\), that spans \(N_B\) frequency channels of the instrument between the signal's lower frequency, \(\nu_L\), and upper frequency, \(\nu_U\), we take 
\begin{equation}
    Z_B(t_n, p) = \frac{1}{\sqrt{N_B}}\sum_{\nu=\nu_L}^{\nu_U}Z(t_n, \nu, p).
    \label{eq:Z_B}
\end{equation}
This sum is precisely constructed so that a clean \(Z_B\) will be standard normal distributed just like a clean \(Z\), so that the same significance threshold can be applied to any sub-band as is applied to a single sample. When RFI in a sub-band is observed for only part of the observation, even a sequence of very weak positive (or negative) outliers at a given time can sum to be greater in absolute value than the desired significance threshold, thereby increasing the sensitivity to RFI that occupies that sub-band. 
We can then iterate similarly as in the single-sample case to flag the observation for sub-band outliers in addition to single-sample outliers, where now we identify the strongest outlier among all suspected RFI sub-bands and times in the observation and then proceed as before. We show an example of the frequency-matched flagger in Figure \ref{fig:Filt_Demo}(b), where it successfully excised DTV RFI belonging to multiple broadcasting channels. A movie showing the status of the observation in each iteration is available in the ancillary files.

The \textsc{ssins} frequency-matched flagger can be adapted to search for any RFI contaminant within the observing band, including RFI that occupies the entire band. In \S\ref{sec:comp}, we illustrate these adaptive capabilities using specific examples of common RFI signals that we observe with the MWA.

\section{Some Common RFI Occupants in MWA Data}

\label{sec:comp}

The sensitivity boost afforded by the baseline averaging allows \textsc{ssins} to identify a wealth of faint RFI that goes undetected by single baseline algorithms. However, RFI that persists throughout the entire observing time can still evade \textsc{ssins} due to the fact that such RFI will always contaminate the mean estimation no matter how deeply it is flagged. We have successfully implemented methods in \textsc{ssins} to compensate for this behavior. Below, we explore the behavior of \textsc{ssins} through examples for three different RFI classifications: broadband streaks, narrowband RFI, and DTV. In each case, the data was passed through \textsc{aoflagger} before making the incoherent noise spectra, except where indicated in \S\ref{sec:NB}.

\subsection{Faint Broadband Streaks}
\label{sec:streaks}

\begin{figure}
    \centering
    \begin{subfigure}[]{
        \includegraphics[width=\columnwidth]{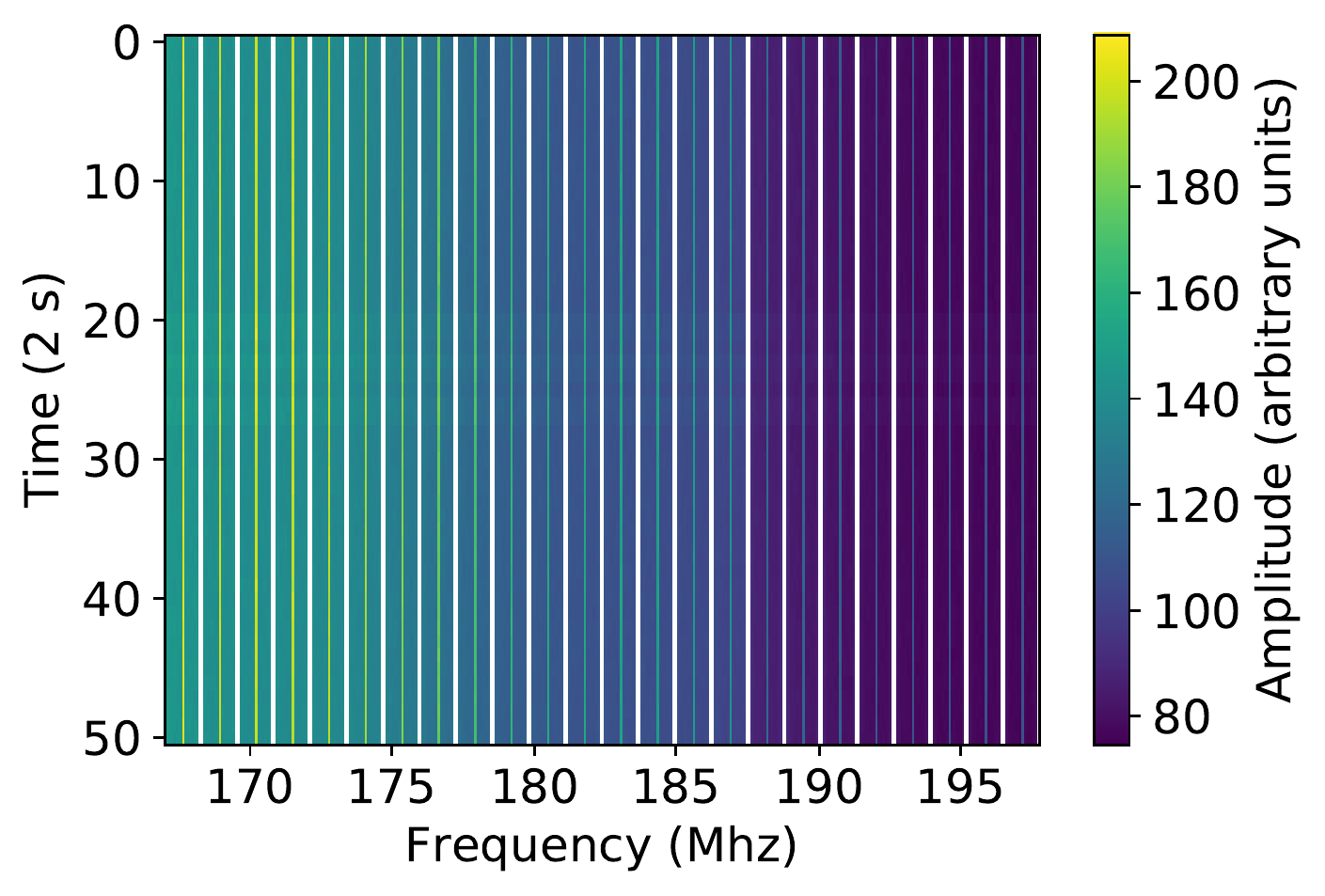}}
   \end{subfigure}
    \begin{subfigure}[]{
        \includegraphics[width=\columnwidth]{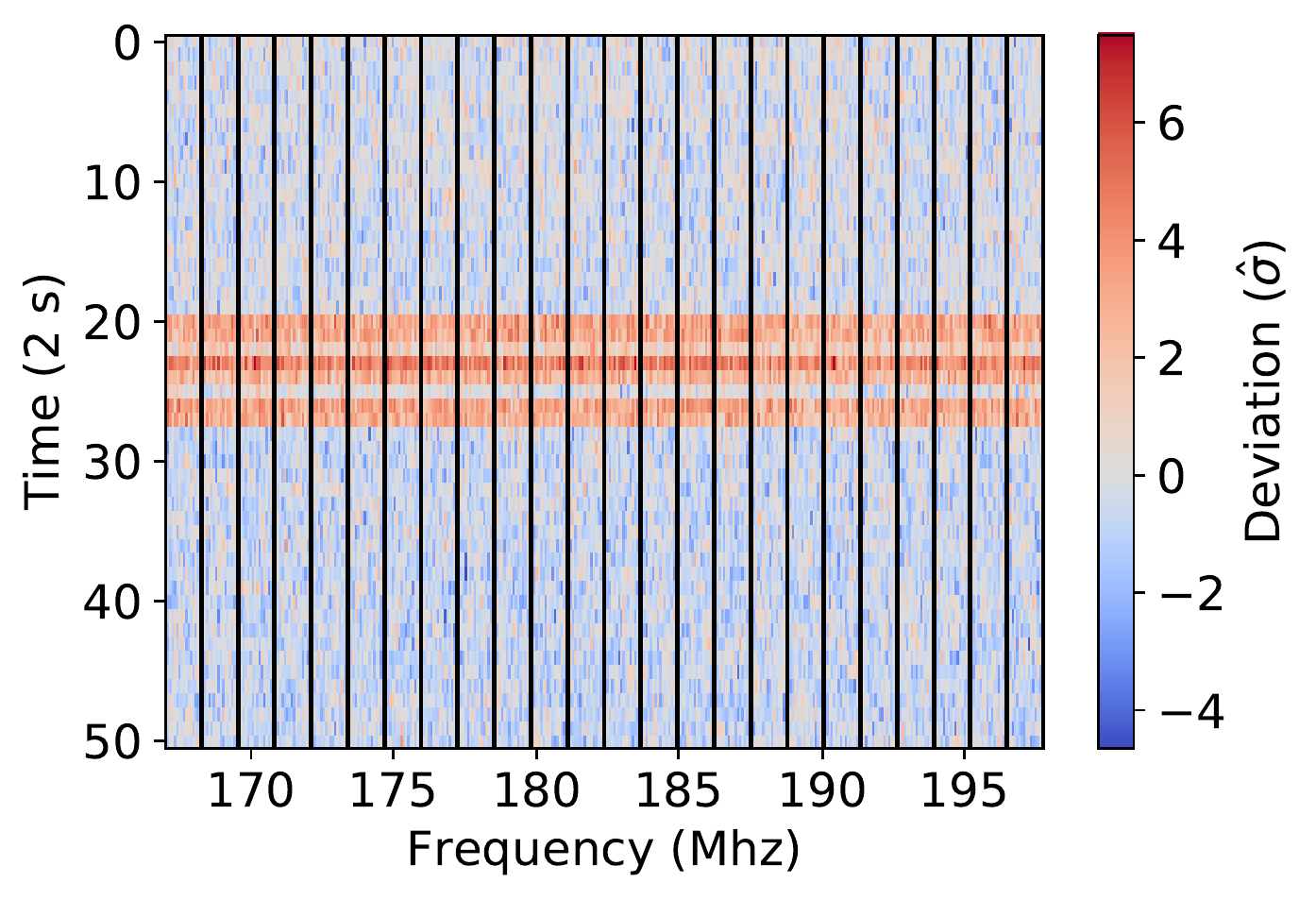}}
    \end{subfigure}
    \begin{subfigure}[]{
        \includegraphics[width=\columnwidth]{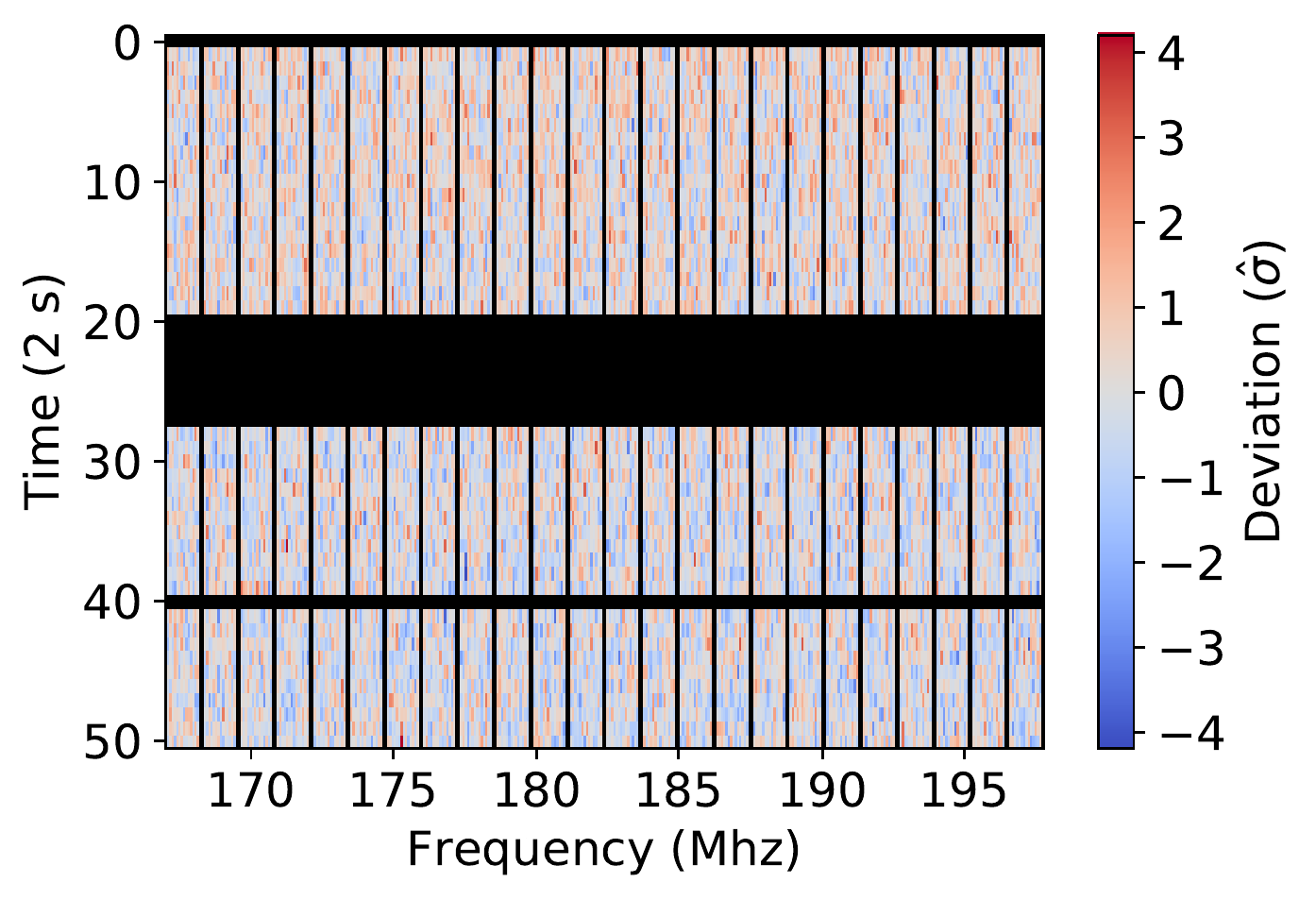}}
    \end{subfigure}
    \caption{The SSINS of an observation which show faint broadband streaks. The pre-existing flags in (a) and (b) are over the coarse band edges, which are routinely flagged in pre-processing due to systematic difficulties. The streaks are barely noticeable to the eye in (a), but the mean-subtracted spectrum in (b) shows them prominently. The frequency-matched flagger excises these streaks neatly, as shown in (c). These exceedingly faint events are often missed by \textsc{aoflagger} and are not uncommon. They can appear isolated within an observation, or in series like this.}
    \label{fig:faint_streaks}
\end{figure}

First, we show an observation with faint streaks that are not band limited. See Figure \ref{fig:faint_streaks}. These streaks are too faint to be detected by the standard single-baseline \textsc{aoflagger} implementation and appear to be quite common in our brief survey of MWA EoR Highband data. The exact nature of these streaks is so far unknown. Since they are not band-limited, we cannot directly appeal to our knowledge of the RFI environment in the same way that we can with the DTV interference shown earlier. It is possible that they are instrumental in origin. Despite not knowing the physical origin of such events, these streaks still occupy a programmable sub-band for the frequency-matched flagger, that is, the whole observing band, and so we can still adequately flag them using the algorithm in this paper.

\subsection{Narrowband Interference}
\label{sec:NB}

\begin{figure}
    \centering
    \begin{subfigure}[]{
        \includegraphics[width=\columnwidth]{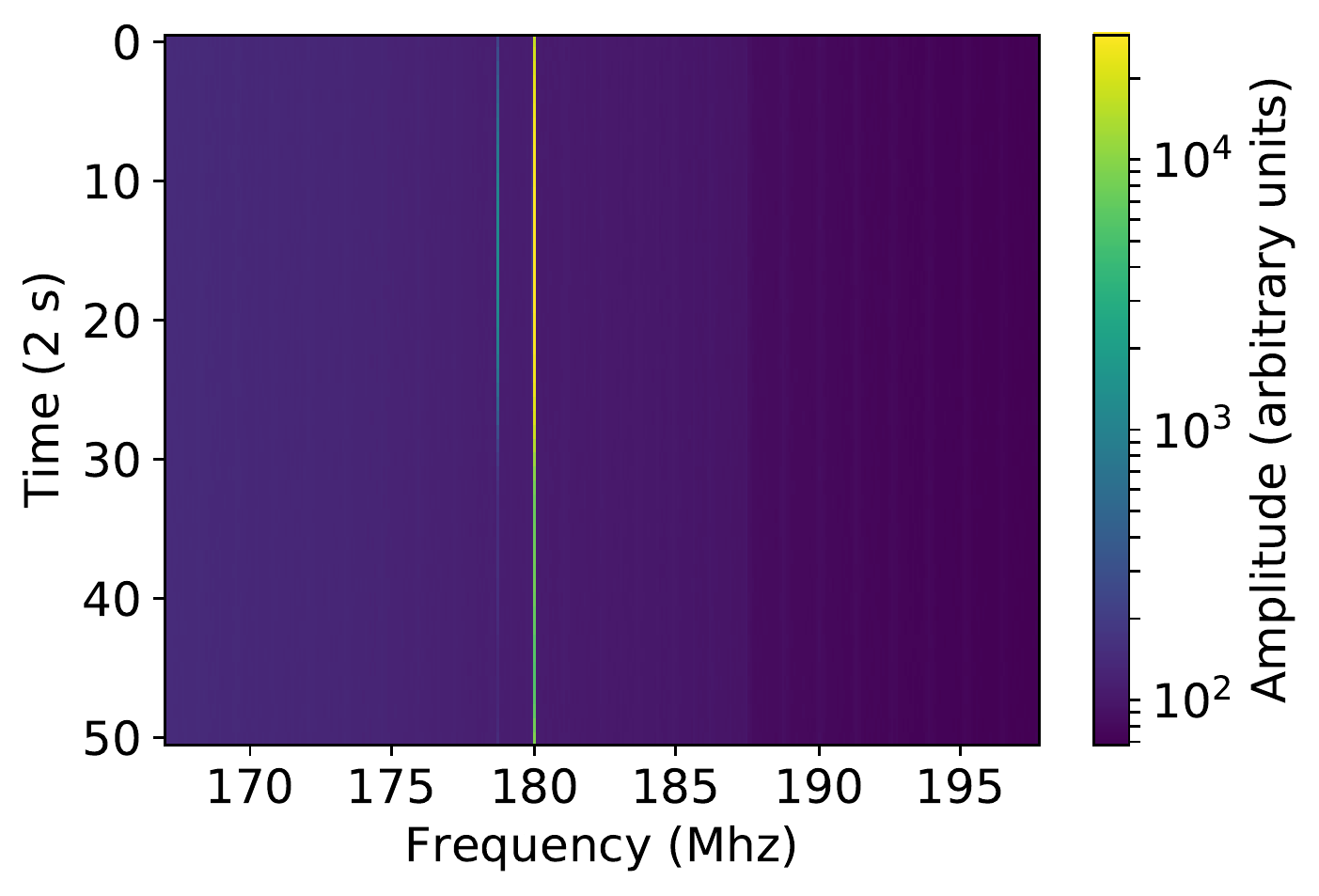}}
   \end{subfigure}
    \begin{subfigure}[]{
        \includegraphics[width=\columnwidth]{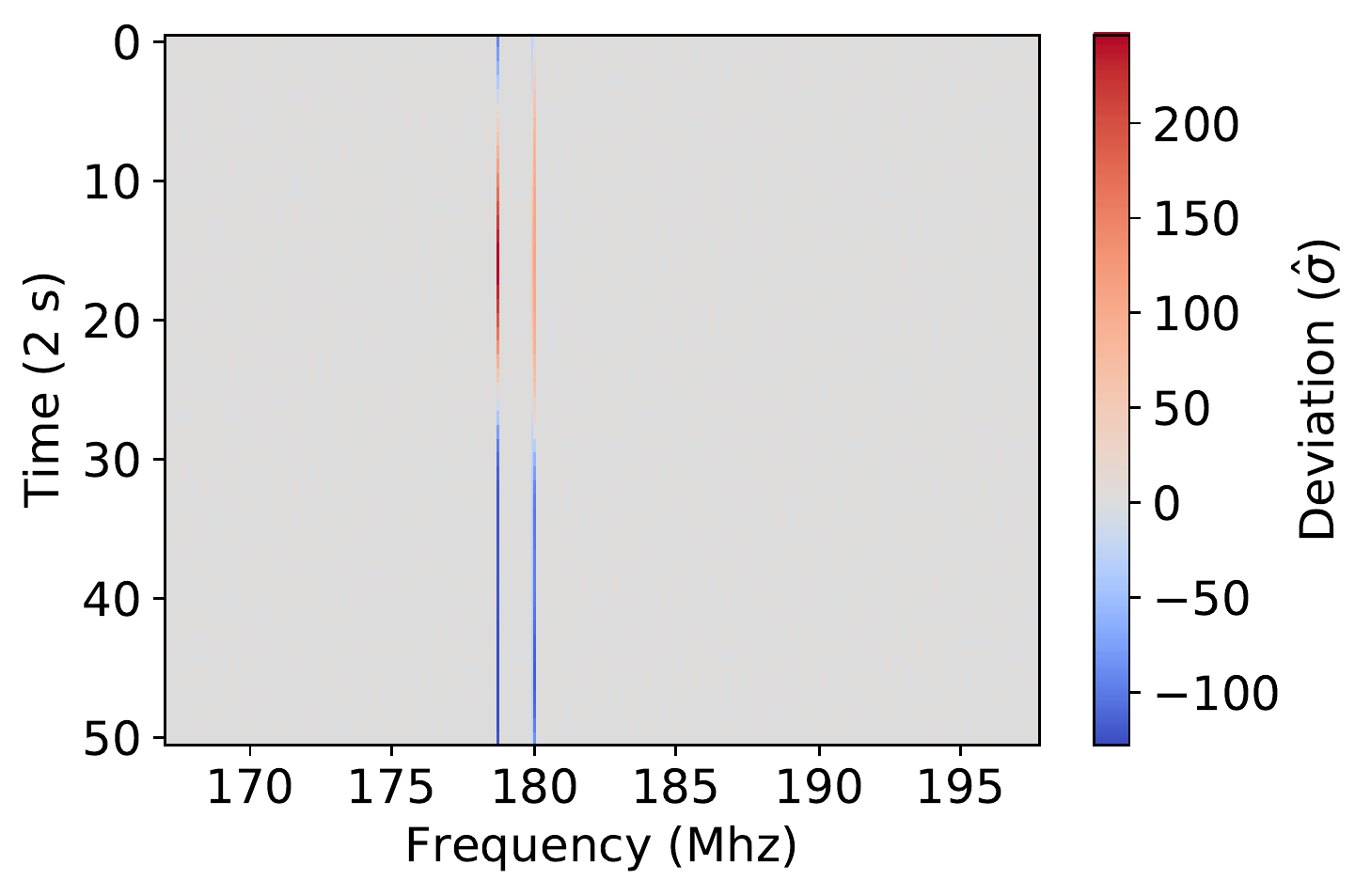}}
    \end{subfigure}
    \begin{subfigure}[]{
        \includegraphics[width=\columnwidth]{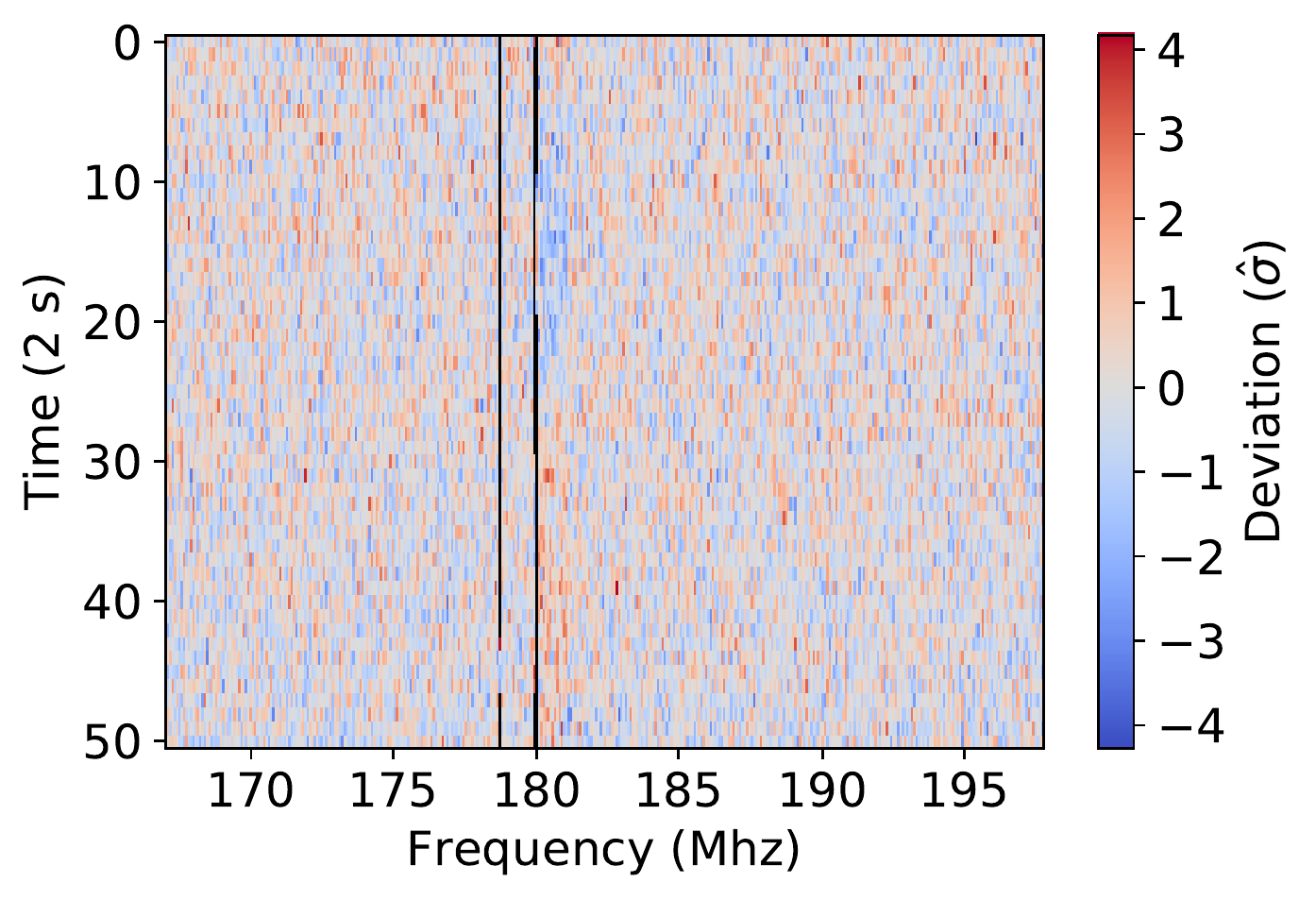}}
    \end{subfigure}
    \caption{Pictured is an example of narrowband RFI that lasts the entirety of an observation. Note the logarithmic color scale in (a). This extremely bright event is a full two orders of magnitude brighter than the typical clean spectrum, and so the finer structure typically seen in an MWA incoherent noise spectrum is washed out in this colormap. We also note a disturbance precisely one coarse channel wide to the right of the main interference event in (c), characterized by a blue trough followed by a red excess. This is the only known instance of this feature in MWA data so far. Its exact nature is unknown. It may be related to the sheer brightness of this RFI event.}
    \label{fig:NB_trouble}
\end{figure}

Next we show an example of narrowband interference observed by the MWA that only occupies one or two fine frequency channels (Figure \ref{fig:NB_trouble}). Though this RFI is present in the observation at dramatically fainter levels after \textsc{aoflagger}, we show the pre-\textsc{aoflagger} spectrum in this case because of the interference's proximity to a coarse band edge, which makes viewing the event difficult after routine flagging of the edges. 

In this case, there are two interference events that are present through the entirety of the observation. The higher frequency line is a full order of magnitude brighter than the lower frequency line, and they are separated by exactly one coarse channel. We hypothesize that the dimmer line is caused by the brighter line's proximity to a coarse band edge, which are prone to aliasing due to the cascaded fourier transform in the digital signal path\footnote{See "Aliases" in \url{https://wiki.mwatelescope.org/display/MP/Memos?preview=/14156367/18481185/MEMO\_CascadedFT\_2012\_05\_25.pdf}}.

\begin{figure*}[t!]
    \centering
    \begin{subfigure}[]{
        \includegraphics[width=0.48\linewidth]{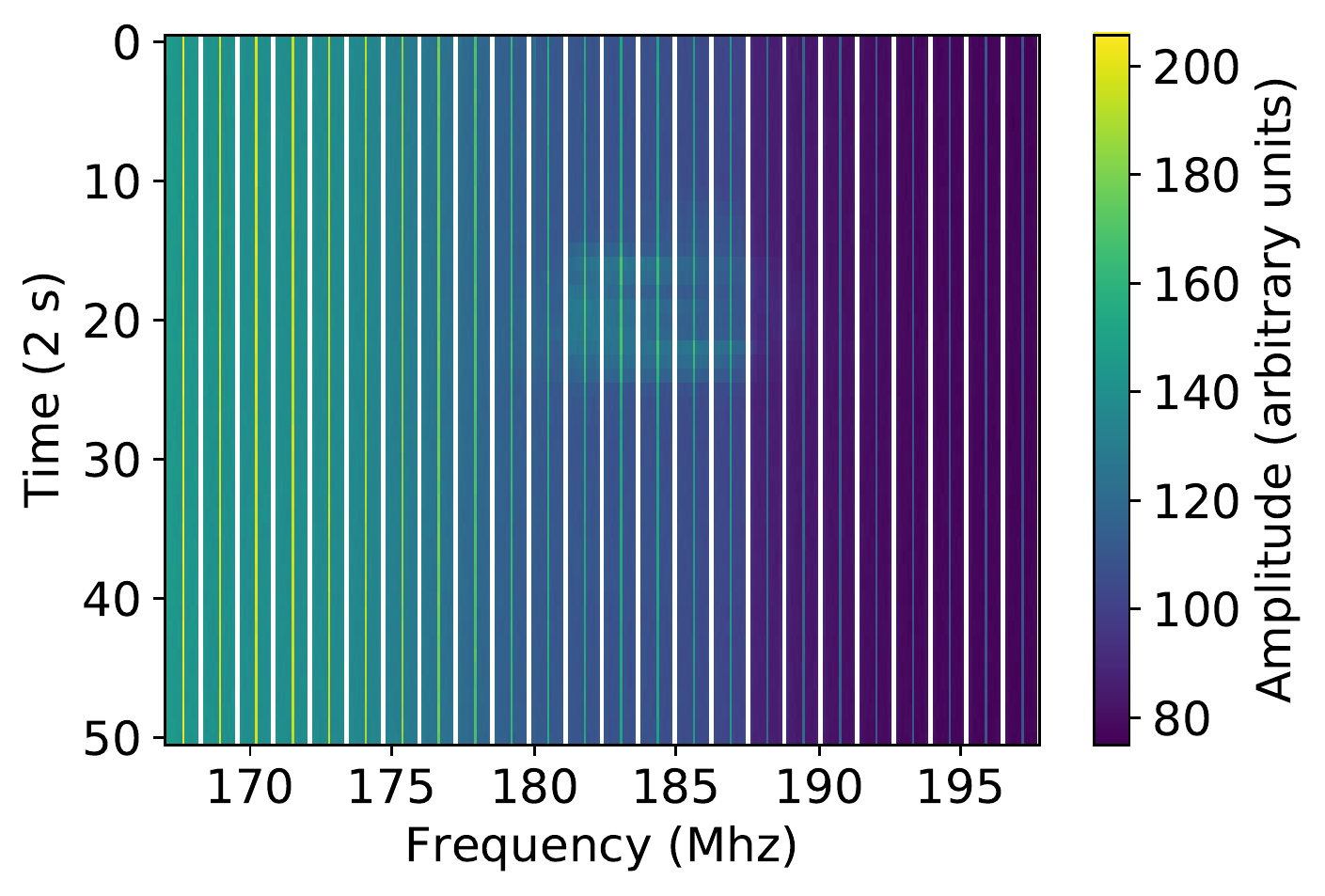}}
    \end{subfigure}
    \begin{subfigure}[]{
        \includegraphics[width=0.48\linewidth]{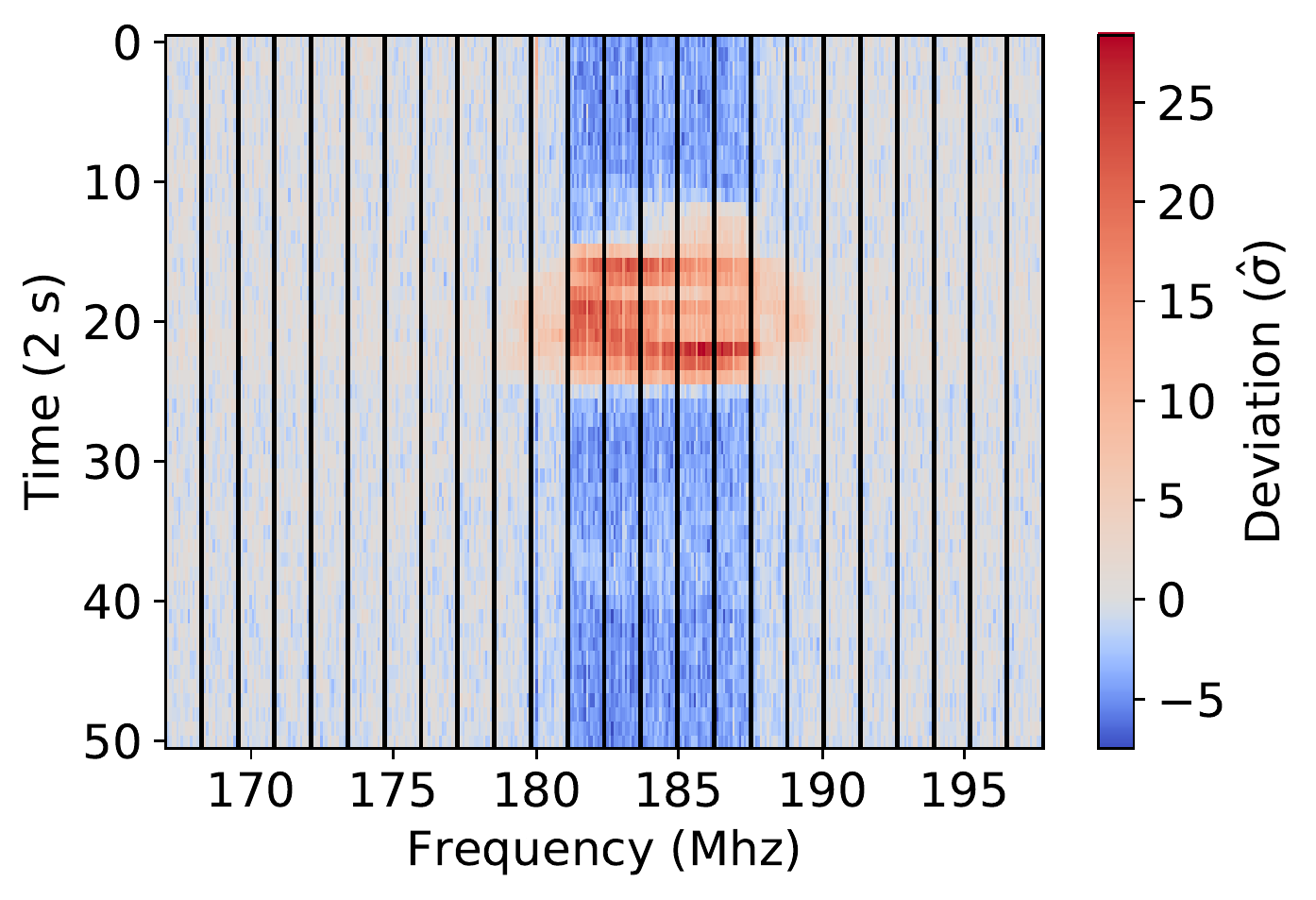}}
    \end{subfigure}
    \begin{subfigure}[]{
        \includegraphics[width=0.48\linewidth]{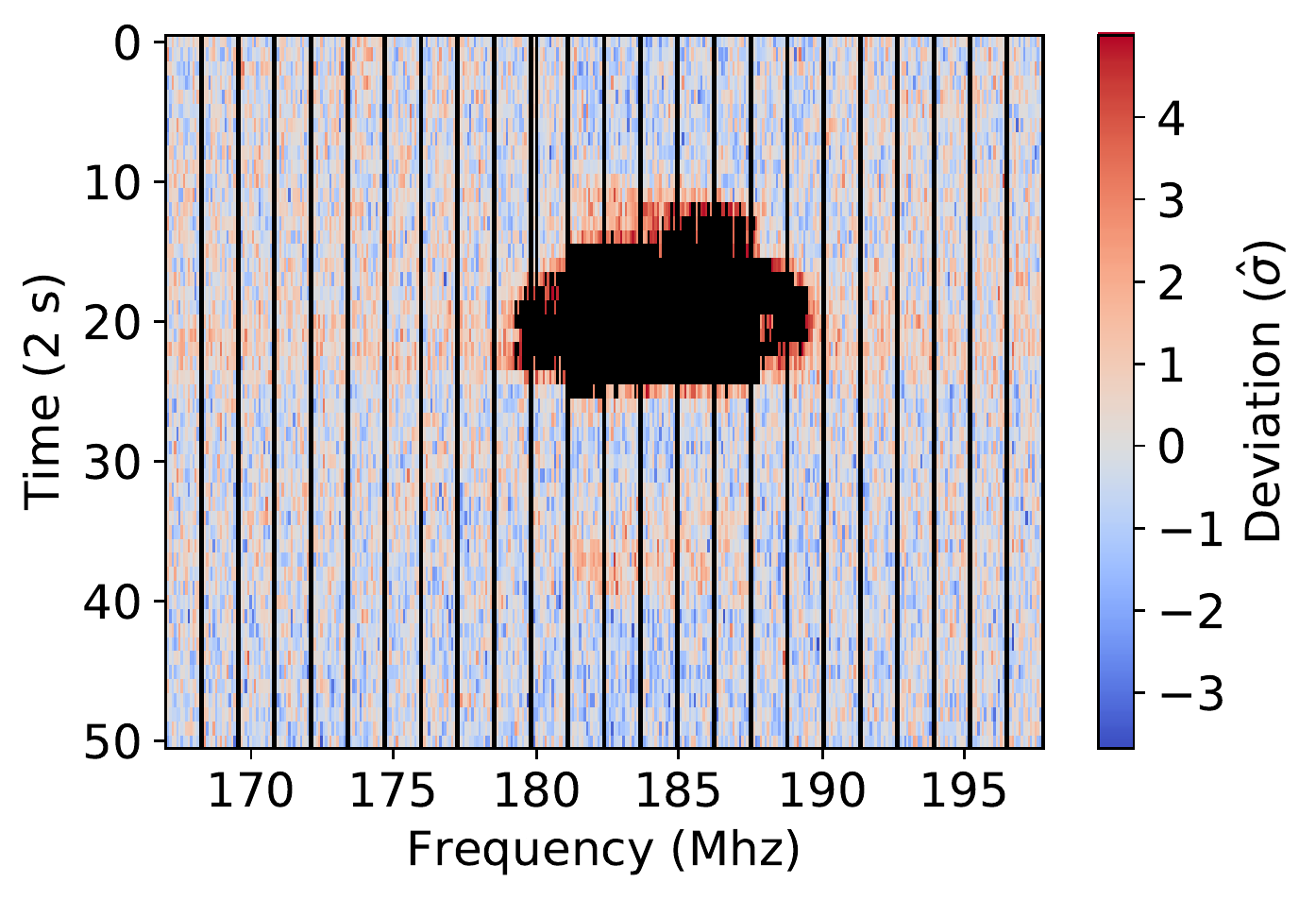}}
    \end{subfigure}[]
    \begin{subfigure}[]{
        \includegraphics[width=0.48\linewidth]{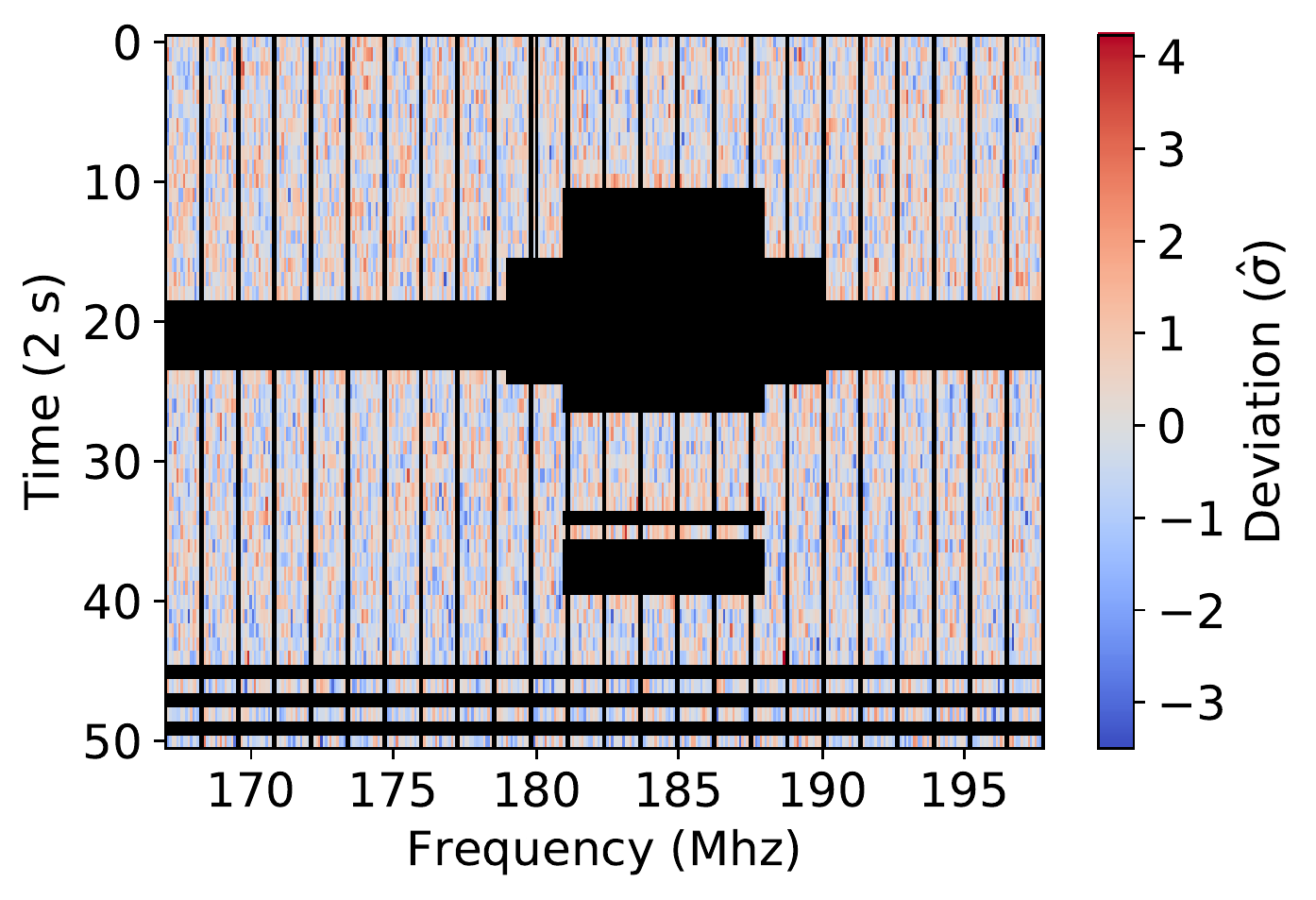}}
    \end{subfigure}
    \caption{The SSINS of a DTV example after applying \textsc{aoflagger} (a), alongside the mean-subtracted spectrm (b), as well as the results of single-sample iterative flagging (c), and frequency-matched flagging (d). From the flags reported by \textsc{aoflagger}, we know it was caught on some baselines, however there is clearly leftover DTV corresponding to DTV channel 7. Several features emerge after single-sample flagging in (c). First, it appears that there is a broadband streak (\S\ref{sec:streaks}) simultaneous with the DTV interference. Second, it appears there may have been a second DTV interference event later in the observation and much fainter. Third, the DTV interference seems to have associated outliers that span more than the advertised 7 Mhz, which is a pre-processing artifact described in the main text.  With the frequency-matched flagger we can search for DTV, broadband streaks, single-sample outliers, and the pre-processing artifact simultaneously. The results are shown in (d), where all notable features are excised.}
    \label{fig:bright7}
\end{figure*}

Narrowband RFI such as this that persists throughout the entire observation poses a detection problem for the \textsc{ssins} algorithm, since the estimation of the thermal parameters will always be contaminated by the RFI. The statistical properties of the RFI usually do not perfectly resemble the thermal noise, but even so, such RFI is almost always incompletely flagged by the \textsc{ssins} frequency-matched flagger. Rarely, there is even RFI whose brightness varies so little over the course of the observation that it is totally camouflaged by the mean-subtraction step. Clearly, narrowband RFI that only occupies some fraction of the observation does not have this problem. In Figure \ref{fig:NB_trouble}(c), we show the mean-subtracted spectrum after frequency-matched flagging. We can see that the fainter event (alias) was flagged almost completely. If we examine the brighter event at the higher coarse channel, we notice the flag mask is thicker at some times compared to others. The RFI seems to occupy two adjacent fine frequency channels. The times indexed 10 to 20 are missing flags in the higher fine frequency channel, while the times indexed 30 to 48 are missing flags in the lower fine frequency channel.

We handle incomplete flagging in the following way. First, note that as we flag more times in a channel, fewer samples enter into the estimation of the mean. Eventually, the uncertainty in the estimator will be so high that the estimate will be untrustworthy. Furthermore, as more and more data within a channel are flagged, the chance that clean data remain in that channel diminishes. Combining these ideas, we set a threshold where once the amount of remaining unflagged data falls below the threshold, the entire remainder of the channel is flagged. The exact value that this threshold should take depends on the RFI environment of the telescope. While many narrowband events in the MWA survey shown in \S\ref{sec:survey} would demand a threshold as aggressive as 0.7 (flag if less than 70\% remains), an extremely brief survey of HERA data shows that a threshold of 0.25 successfully flags many of HERA's persistent narrowband occupants.

\subsection{A DTV Signal}
\label{sec:bright_TV}

Finally, we show an observation with DTV interference that was partially caught by \textsc{aoflagger}. In this case, the DTV was caught on some baselines, but not all of them, so we can see a leftover DTV footprint in the noise spectrum. See Figure \ref{fig:bright7}.

We perform a similar demonstration as with the fainter DTV events shown in \S\ref{sec:Match-Shape} by first flagging only bright single-sample outliers and then checking to see if fainter occupants lie beneath the single-sample threshold (Figure \ref{fig:bright7}(c)). A notable feature that emerges from single-sample flagging is that the cluster of significant outliers around the DTV event seem to span more than 7 Mhz. The excess width of this event is not seen in the pre-\textsc{aoflagger} spectrum (not shown), indicating that this is a pre-processing artifact. As described in ~\citet{O2012}, \textsc{aoflagger} deploys a morphological detection algorithm that can overflag broad contaminants in a baseline by an amount proportional to the algorithm's user-set aggression threshold. Indeed, summing the \textsc{aoflagger} flags for this observation over the set of baselines does show overflagging of this event in frequency by an amount similar to the feature we see in the incoherent noise spectrum. A feature of this size manifests in the incoherent noise spectrum due to the fact that visibilities have been averaged in time and frequency relative to the operating time-frequency resolution of \textsc{aoflagger}. Flags are applied before averaging, so time-frequency bins with fewer samples entering them will have noise that has not been averaged down as much compared to those bins in which all possible contributing samples were averaged together. The overflagged bins contributing to the incoherent noise spectrum are then brighter than the surrounding uncontaminated ones, thus appearing like RFI to our statistical test. We can adapt the frequency-matched flagger with a custom sub-band to identify this pre-processing artifact. The results of frequency-matched flagging are shown in Figure \ref{fig:bright7}(d).

\begin{figure*}[t!]
    \centering
    \includegraphics[width=\textwidth]{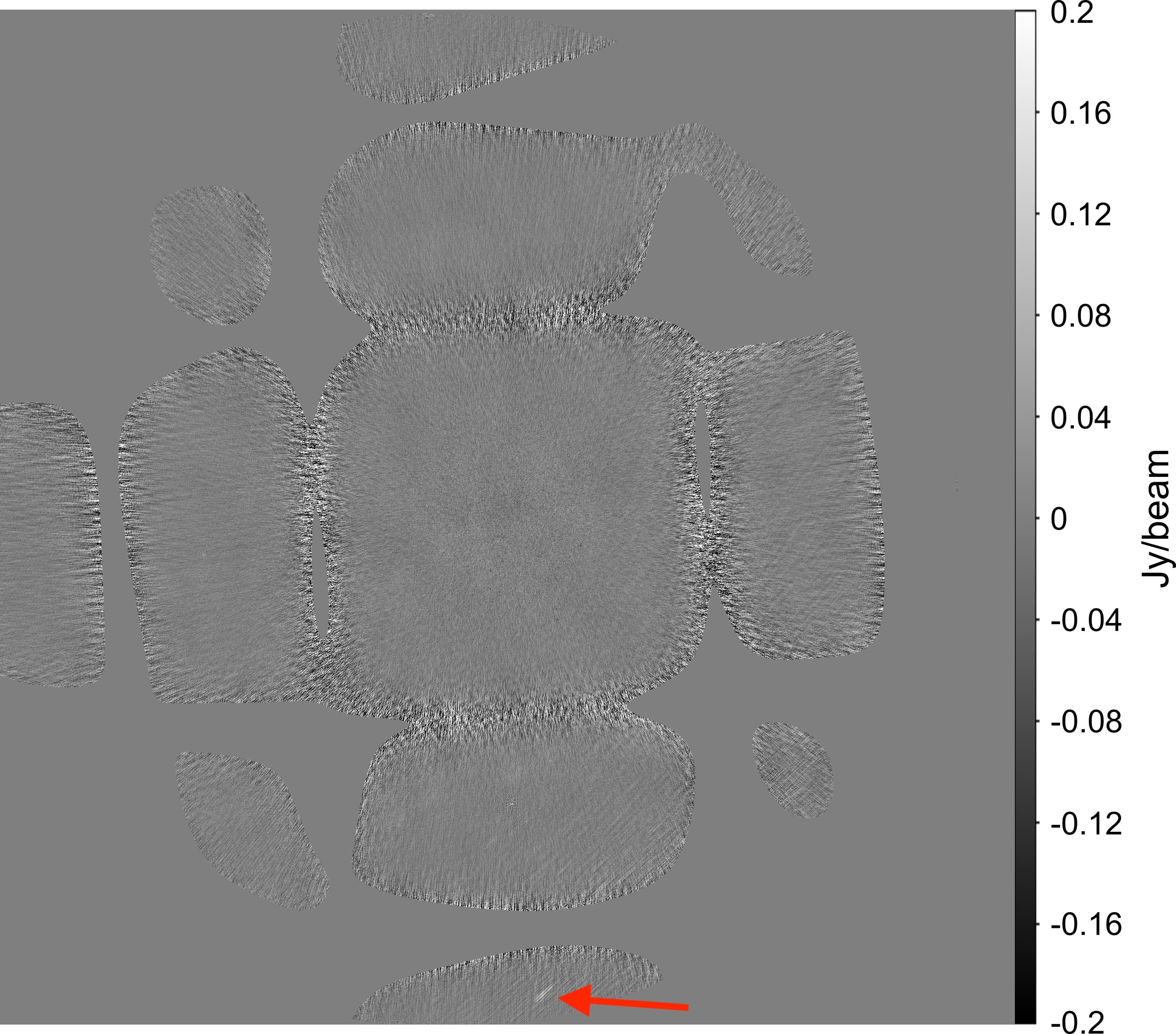}
    \caption{A full-sky horizon-to-horizon image of the DTV channel 6 event in Figure \ref{fig:MS_SSINS}(b), with 52000 GLEAM sources removed using FHD. Other than the shape of the MWA beam and some diffuse structure not included in the calibration/subtraction model, the image is largely featureless. However, all the way in the second southern sidelobe lies a faint streak belonging to the DTV6 event (annotated with a bright red arrow). We hypothesize that the DTV signal is reflecting off of an aircraft on the southern horizon into the array. This is but one example of a collection of observations that show a similar feature in the second southern sidelobe, some of which are at the same time of night but on different dates, as one might expect from scheduled flights to and from Perth.}
    \label{fig:TV6_sidelobe}
\end{figure*}

This type of DTV interference is extremely common in MWA EoR Highband Observations. Roughly one third of the observations included in the EoR limit in \citet{Beardsley2016} had some trace of DTV RFI according to the frequency-matched flagger. These observations were subsequently removed in a reduction of the same data set in \citet{Barry2019b}, as discussed in \S\ref{sec:survey}.  While in this case \textsc{aoflagger} was able to remove some of the interference, it is often the case that the contamination is beneath the level where \textsc{aoflagger} can make an appreciable difference.

\section{Imaging DTV Interference}
\label{sec:images}

The MWA's extremely remote location makes it unlikely that it would directly observe a DTV transmission. However, it is clear from noise spectra such as those in the previous sections that the MWA is observing DTV. There are several hypotheses to explain this, including tropospheric ducting and reflection off of aircraft or satellites. We explore these possibilities by imaging DTV events found using \textsc{ssins}.

\begin{figure*}[t!]
    \centering
    \includegraphics[width=\textwidth]{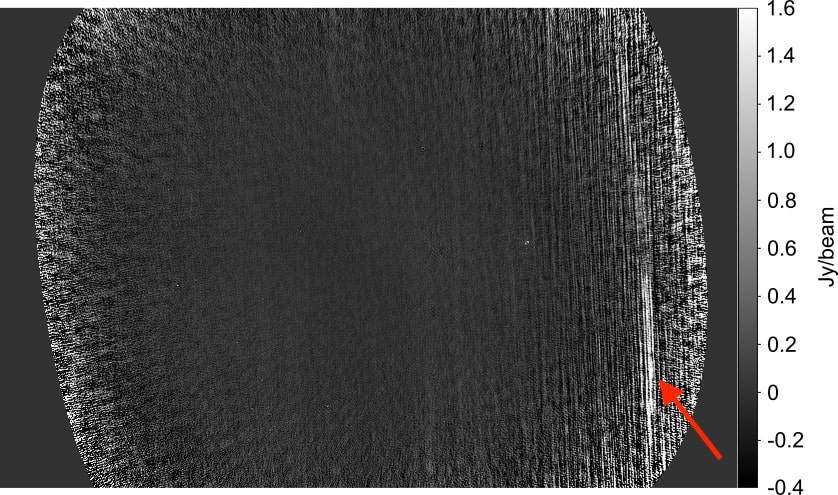}
    \caption{The Stokes I residual image of the DTV event in Figure \ref{fig:bright7}, showing just the primary beam (center lobe in Figure \ref{fig:TV6_sidelobe}). This is 34 seconds of data. The broad North-South streak in the Eastern edge of the primary beam (annotated with a bright red arrow) suggests possible motion of a source - likely a reflective aircraft. A snapshot-by-snapshot movie is available in the ancillary files.}
    \label{fig:bright7_image}
\end{figure*}

The images were made using the Fast Holographic Deconvolution (FHD) software\footnote{\url{https://github.com/EoRImaging/FHD}} ~\citep{Sullivan2012, Barry2019a}. So that RFI could not drive the calibration solution, we calibrated on a clean part of each observation and transferred that calibration to the contaminated part of the respective observation. In order to see these faint DTV events in the images, it was necessary to subtract out 52000 sources from the GaLactic and Extragalactic All-sky MWA (GLEAM) catalog \citep{Hurley-Walker2017} using FHD. Example images using the observations from Figures \ref{fig:MS_SSINS}(b) and \ref{fig:bright7} are shown in Figures \ref{fig:TV6_sidelobe} and \ref{fig:bright7_image}, respectively. We separated Figure \ref{fig:bright7_image} into 2s-snapshots and made a movie (available in the ancillary files), where it is clear that some sort of flying object moving nearly due North-South is reflecting DTV into the array.

 Due to the nonzero extent of the moving source, we can rule out the possibility of satellites and near-Earth asteroids, which would appear point-like. Aircraft are reflective and fly low enough to appear extended in the image. The speed (extent) of the object is atypically slow (large) for a commercial jet such as a Boeing 737 flying at a typical cruising altitude of 10 km. A parallactic estimation of the object's altitude\footnote{See \cite{Loi2016} and citations therein for other uses of this type of altitude estimation} ultimately proved inconclusive due to the faintness of the RFI (Xiang Zhang, personal communication). As an alternative candidate, we suggest this could be a slow, low-flying bush plane whose apparent angular size may be due to its low height and near-field effects.

\section{Frequency-Matched Flagger Customization}
\label{sec:custom}

In this section, we describe the process of developing a customized \textsc{ssins} frequency-matched flagger for a radio telescope. This is an exploratory process wherein the user becomes familiar with the incoherent noise spectra of their telescope. Some of the spectra features will be due to RFI, while other features may point to subtleties of the instrument. Once equipped with a thorough catalog of sub-bands to search, the user may process large amounts of data in an automated way. To exemplify what can be attained from this process, we summarize the RFI occupancy analysis for the data used in the EoR limit presented in \citet{Barry2019b}.

\subsection{Exploring the Data and Developing a Frequency-Matched Flagger}
\label{sec:howto}

The first step to applying \textsc{ssins} to a dataset is acquiring the software. It is implemented in \textsc{python} along with a comprehensive set of unit tests. It is publicly available on GitHub\footnote{\url{https://github.com/mwilensky768/SSINS}}, where the user will also find installation instructions including the list of dependencies: \textsc{pyuvdata}\footnote{\url{https://github.com/RadioAstronomySoftwareGroup/pyuvdata}}, \textsc{numpy}, \textsc{scipy}, \textsc{six}, \textsc{h5py}, \textsc{pyyaml}, \textsc{astropy}, and optionally \textsc{matplotlib}. There are tutorials and other documentation available\footnote{\url{https://ssins.readthedocs.io/en/latest}} with simple usage examples for getting started.

After becoming familiar with the software, the next step is to examine the data, typically by hand, without attempting to use the frequency-matched flagger. While going through a survey of observations baseline by baseline in a modern radio telescope would be untenable, \textsc{ssins} compresses the data of each observation into two dynamic spectra per polarization: a raw incoherent noise spectrum and its mean-subtracted form. It is useful to look at both forms of the spectra side-by-side, particularly if narrow RFI is expected. In this way, an entire season of observations can be examined in short order.

While features in the incoherent noise spectra may correspond well with known factors in the RFI environment, it is also likely that \textsc{ssins} will reveal RFI and even other instrumental features that had not been previously observed or considered. As the user catalogs the various characteristics of the data, they can build a dictionary of different sub-bands or single frequencies that appear commonly occupied. These occupations, regardless of their physical cause, can be input into the frequency-matched flagger settings. For example, some faint features in HERA incoherent noise spectra were later found to correspond with correlator malfunctions, some of which had been previously identified in other cases through other means. While technically this was not RFI per se, it was a feature that could be identified and removed with \textsc{ssins}.

Once the set of identifiable contaminants has been collected, the frequency-matched flagger can be deployed for a first round of flagging. The frequency-matched flagger is very quick. A 2-minute, 30 Mhz MWA spectrum can be flagged in less than a second, while a 10-minute, 100 Mhz HERA spectrum that is significantly more contaminated typically takes between 20 and 60 seconds. As with the initial inspection, the results of the frequency-matched flagger can be quickly ascertained by eye. Oftentimes, fainter RFI than could be initially seen in the noise spectra is unearthed as the frequency-matched flagger iterates. Of course, the user will wish to add these to the dictionary of occupants. If no more identifiable sub-band occupancies are revealed, one may wish to adjust the parameters of the flagger, such as the significance threshold. This iterative process of data examination and deep cleaning is often illuminating.

\subsection{Results with a Season of MWA Data}
\label{sec:survey}

\begin{figure*}[t]
    \centering
    \begin{subfigure}[]{
        \includegraphics[width=0.48\linewidth]{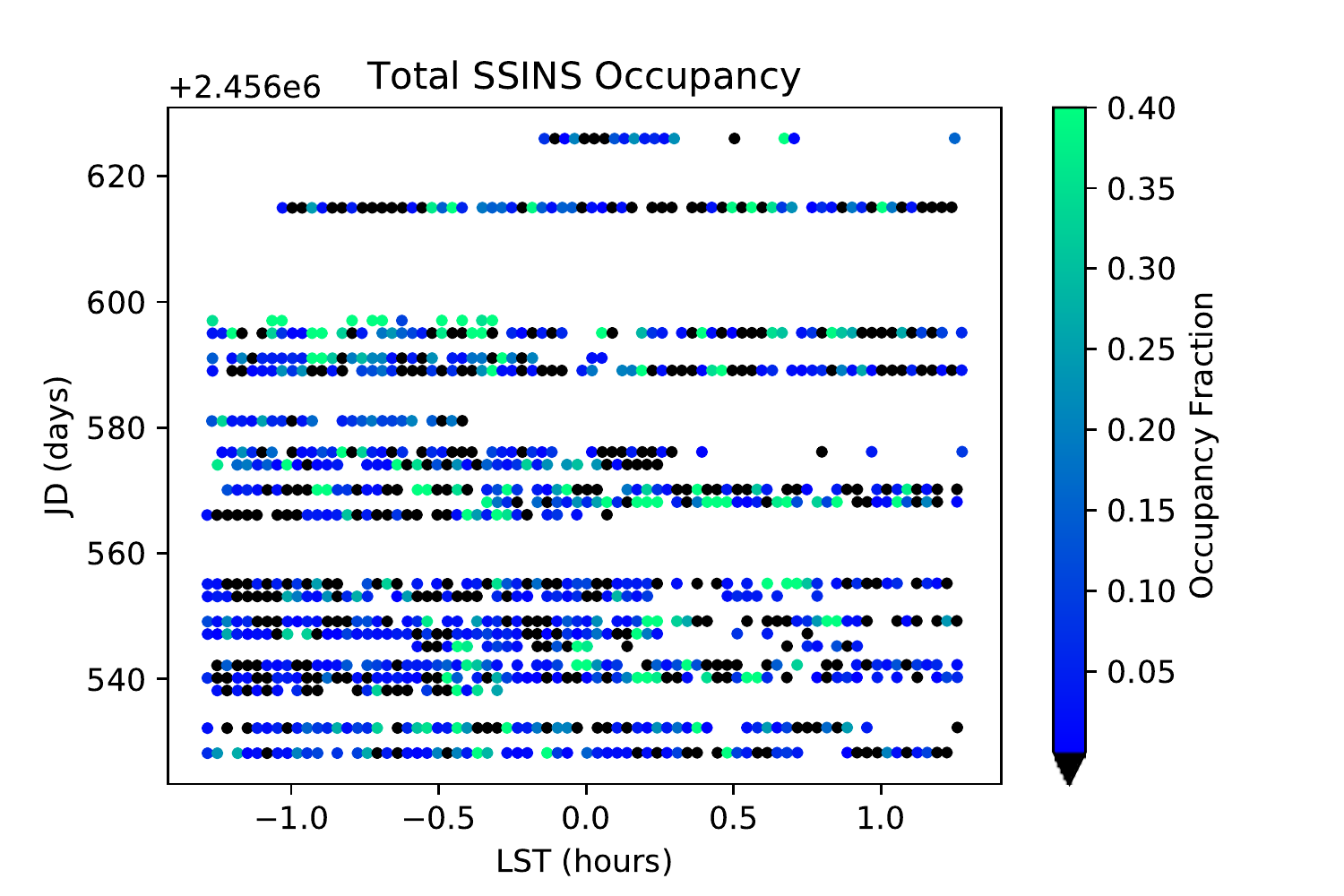}}
    \end{subfigure}
    \begin{subfigure}[]{
        \includegraphics[width=0.48\linewidth]{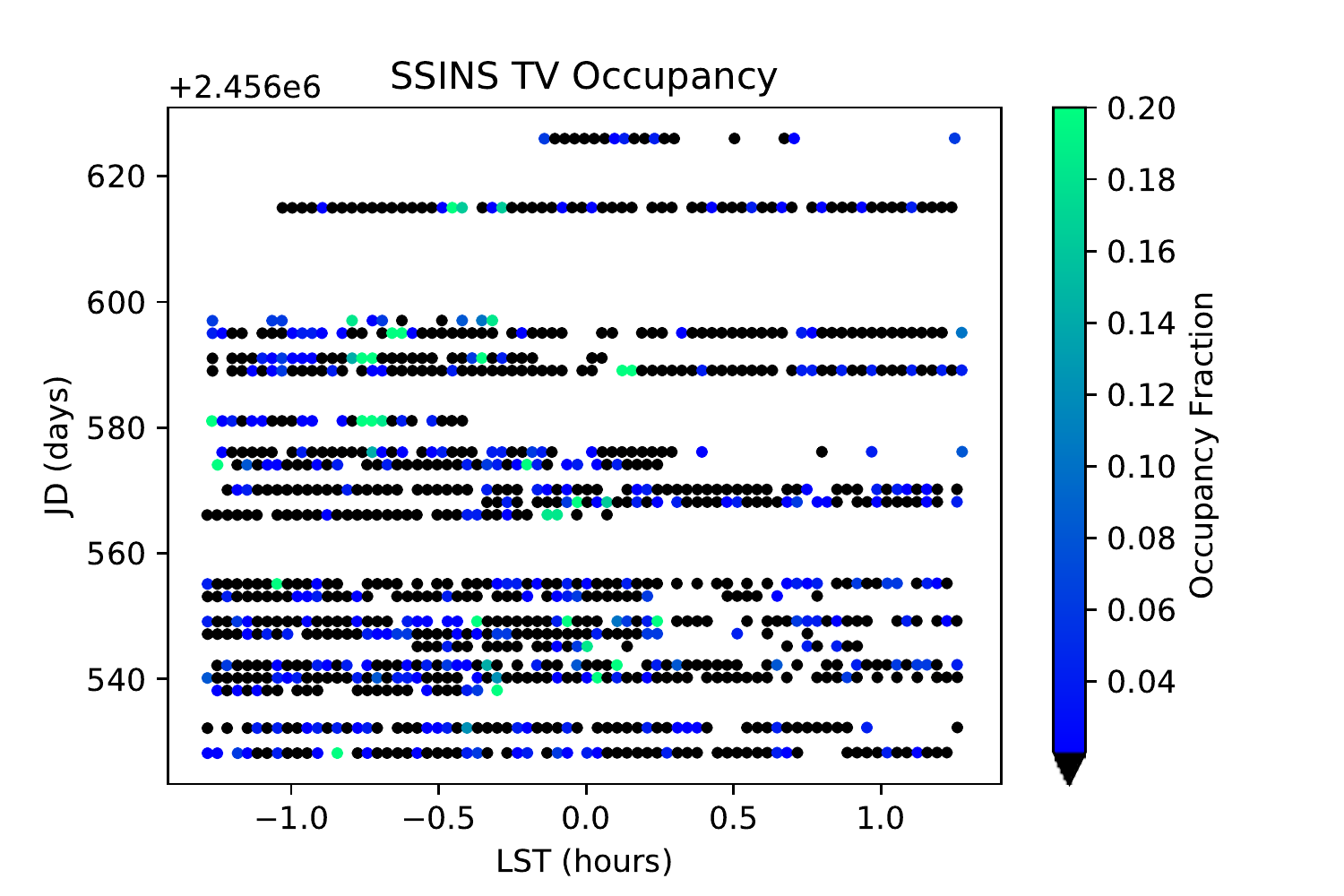}}
    \end{subfigure}
    \caption{(a) A scatter plot of the total RFI occupation in a season of data as seen by \textsc{ssins}. (b) A scatter plot of the DTV occupancy as seen by \textsc{ssins}. In each panel, each circle represents a single two-minute observation. Each line of circles is a night of observations on the Julian Date (JD) shown on the vertical axis, while the horizontal axis gives that observation's local sidereal time (LST) in hours. Black circles were not found to have additional RFI after \textsc{aoflagger}. All observations outside of the plotted LST range and all missing circles in a night are observations that were removed by a previous jackknife test, detailed in \cite{Beardsley2016}. In (a), the color shows the fraction of samples in the noise spectra found to be contaminated, disregarding coarse band edges, while in (b), the color shows the fraction of times contaminated by DTV interference, regardless of which broadcasting channels were identified. These plots provide an occupancy overview of the season, letting one pick out particularly bad days or times of night by eye, such as the line of high occupancy observations nearest JD 2456600, which was a day that also had many observations removed by the previous jackknife test.}
    \label{fig:RFI_occ_scatter}
\end{figure*}

The \textsc{ssins} package itself was developed and tested in the manner described above using a season's worth of data from 2013. This same dataset was used for the EoR power spectrum limit featured in \citet{Beardsley2016} as well as Barry et al. (in revidew). Below, we describe the filter settings for an RFI analysis of this dataset and also summarize occupancy levels.

The data in this analysis was pre-processed with \textsc{cotter}, which uses \textsc{aoflagger} to identify and flag bright RFI. Then, \textsc{ssins} was run on the data in order to identify and catalog leftover faint RFI. We used a significance threshold of 5 and sought single sample outliers, broadband streaks, DTV interference, as well as the broader DTV sub-band discussed in \S\ref{sec:bright_TV}, all of which are occupants detailed in \S\ref{sec:comp}. We also completely flagged a fine frequency channel if it ever reached an occupancy fraction of 0.7 during frequency-matched flagging. We present RFI occupancy for the season in Figure \ref{fig:RFI_occ_scatter}, summarizing total RFI occupancy as seen by \textsc{ssins} after \textsc{aoflagger} as well as DTV occupancy. Such figures can be used to pick out particularly bad days from the season and search for patterns in time.

The occupancy data from \textsc{ssins} was used to make data cuts for the sake of improving the EoR limit in \citet{Barry2019b}. We did not reflag the data using \textsc{ssins}, but instead just used its outputs to develop quality metrics for observations. Ultimately, we cut all observations with any trace of DTV as well as all observations with greater than 40\% occupancy. We found that this improved the limit despite removing roughly 1/3 of the observations originally included. 

\section{Discussions and Conclusions}
\label{sec:conc}

We have described the \textsc{ssins} RFI detection algorithm in detail. The substantial sensitivity boost afforded by the incoherent average over the baselines allows for detection of faint RFI that escapes other high-performing single-baseline algorithms. We demonstrated its effectiveness on several different types of RFI found in the MWA EoR highband, including DTV interference. We have implemented countermeasures for persistent RFI with some success, but we plan to make improvements to these detection efforts. Overall, we observe that the increased sensitivity afforded by \textsc{ssins} helps us understand the general pervasiveness of faint RFI.

Developing a custom \textsc{ssins} frequency-matched flagger for a new RFI environment is an iterative process that is often quite instructive. We confirmed the quality of our own EoR highband frequency-matched flagger by successfully imaging faint DTV reflected from aircraft as well as successfully improving the EoR limit in \citet{Barry2019b}. To ease the development process, we have written tutorials for basic software usage and provided an overview of how \textsc{ssins} can be used in practice.

Though \textsc{ssins} is capable of detecting fainter RFI than current state-of-the-art algorithms and is a general improvement to the field, we plan to further improve the methodology to exploit its full capabilities. We propose the following items as future work:
\begin{itemize}
    \item \textbf{Improved Narrowband RFI Detection}: The current implementation tends to underflag narrowband RFI that persists through the entire observation. An algorithm that preliminarily identifies entirely contaminated channels and interpolates over these channels during mean estimation will allow for more complete flagging.
    
    \item \textbf{Blind Sub-band Hunting}: Presently, the user must input specific sub-bands into the frequency-matched flagger that are established from knowledge of the RFI environment or a hand grading of the data. An algorithm that searches for likely occupied sub-band candidates would ease analysis of extremely large data sets.
    
    \item \textbf{EoR RFI Power Spectrum Shape Characterization}: The effect of faint RFI on an EoR power spectrum has not yet been precisely characterized. We aim to theoretically predict particular RFI footprints in the power spectrum and then subsequently confirm the presence of these footprints in power spectra made from MWA data.
    
    \item \textbf{Data Retention for EoR Limit}: In \citet{Barry2019b}, we used \textsc{ssins} to entirely remove poor-quality observations from the limit calculation, rather than propagating flags from the frequency-matched flagger to the visibilities and keeping any remaining unflagged data within these observations. We expect the EoR limit analysis could be more sensitive if we include remaining clean data from previously cut observations.
\end{itemize}
As \textsc{ssins} is applied to more radio projects in the future, more possibilities for enhancements will be found. In this light, we look forward to exploring the full capabilities of the \textsc{ssins} framework.

\section{Acknowledgements}

MW would like to acknowledge Xiang Zhang for helpful discussions. We would like to thank the development teams of \textsc{pyuvdata}, \textsc{numpy}, \textsc{scipy}, \textsc{six}, \textsc{h5py}, \textsc{pyyaml}, \textsc{astropy}, and \textsc{matplotlib}, which enabled this work. This work was directly supported by NSF grants AST-1643011, AST-1613855, and OAC-1835421. Computation on the Amazon Web Services public cloud was supported by the University of Washington student-led Research Computing Club with funding provided by the University of Washington Student Technology Fee Committee. NB is supported by the Australian Research Council Centre of Excellence for All Sky Astrophysics in 3 Dimensions (ASTRO 3D), through project number CE170100013. This scientific work makes use of the Murchison Radio- astronomy Observatory, operated by CSIRO. We acknowledge the Wajarri Yamatji people as the traditional owners of the Observatory site. Support for the operation of the MWA is provided by the Australian Government (NCRIS), under a contract to Curtin University administered by Astronomy Australia Limited. We acknowledge the Pawsey Supercomputing Centre which is supported by the Western Australian and Australian Governments.

\bibliography{library}

\begin{thebibliography}{}
\expandafter\ifx\csname natexlab\endcsname\relax\def\natexlab#1{#1}\fi
\providecommand{\url}[1]{\href{#1}{#1}}
\providecommand{\dodoi}[1]{doi:~\href{http://doi.org/#1}{\nolinkurl{#1}}}
\providecommand{\doeprint}[1]{\href{http://ascl.net/#1}{\nolinkurl{http://ascl.net/#1}}}
\providecommand{\doarXiv}[1]{\href{https://arxiv.org/abs/#1}{\nolinkurl{https://arxiv.org/abs/#1}}}

\bibitem[{{An} {et~al.}(2017){An}, {Chen}, {Mohan}, \& {Lao}}]{An2017}
{An}, T., {Chen}, X., {Mohan}, P., \& {Lao}, B.~Q. 2017, Acta Astronomica
  Sinica, 58, 43.
\newblock \doarXiv{1711.01978}

\bibitem[{Athreya(2009)}]{Athreya_2009}
Athreya, R. 2009, The Astrophysical Journal, 696, 885,
  \dodoi{10.1088/0004-637x/696/1/885}

\bibitem[{{Baan}(2019)}]{Baan2019}
{Baan}, W.~A. 2019, Journal of Astronomical Instrumentation, 8, 1940010,
  \dodoi{10.1142/S2251171719400105}

\bibitem[{{Barry}(2018)}]{Barry2018}
{Barry}, N. 2018, PhD thesis, University of Washington

\bibitem[{Barry {et~al.}(2019{\natexlab{a}})Barry, Beardsley, Byrne, Hazelton,
  Morales, Pober, \& Sullivan}]{Barry2019a}
Barry, N., Beardsley, A.~P., Byrne, R., {et~al.} 2019{\natexlab{a}},
  Publications of the Astronomical Society of Australia, 36, e026,
  \dodoi{10.1017/pasa.2019.21}

\bibitem[{Barry {et~al.}(2019{\natexlab{b}})Barry, Wilensky, Trott, Pindor,
  Beardsley, Hazelton, Sullivan, Morales, Pober, Line, Greig, Byrne, Lanman,
  Li, Jordan, Joseph, McKinley, Rahimi, Yoshiura, Bowman, Gaensler, Hewitt,
  Jacobs, Mitchell, Shankar, Sethi, Subrahmanyan, Tingay, Webster, \&
  Wyithe}]{Barry2019b}
Barry, N., Wilensky, M., Trott, C.~M., {et~al.} 2019{\natexlab{b}}, The
  Astrophysical Journal, 884, 1, \dodoi{10.3847/1538-4357/ab40a8}

\bibitem[{{Beardsley} {et~al.}(2016){Beardsley}, {Hazelton}, {Sullivan},
  {Carroll}, {Barry}, {Rahimi}, {Pindor}, {Trott}, {Line}, {Jacobs}, {Morales},
  {Pober}, {Bernardi}, {Bowman}, {Busch}, {Briggs}, {Cappallo}, {Corey}, {de
  Oliveira-Costa}, {Dillon}, {Emrich}, {Ewall-Wice}, {Feng}, {Gaensler},
  {Goeke}, {Greenhill}, {Hewitt}, {Hurley-Walker}, {Johnston-Hollitt},
  {Kaplan}, {Kasper}, {Kim}, {Kratzenberg}, {Lenc}, {Loeb}, {Lonsdale},
  {Lynch}, {McKinley}, {McWhirter}, {Mitchell}, {Morgan}, {Neben},
  {Thyagarajan}, {Oberoi}, {Offringa}, {Ord}, {Paul}, {Prabu}, {Procopio},
  {Riding}, {Rogers}, {Roshi}, {Udaya Shankar}, {Sethi}, {Srivani},
  {Subrahmanyan}, {Tegmark}, {Tingay}, {Waterson}, {Wayth}, {Webster},
  {Whitney}, {Williams}, {Williams}, {Wu}, \& {Wyithe}}]{Beardsley2016}
{Beardsley}, A.~P., {Hazelton}, B.~J., {Sullivan}, I.~S., {et~al.} 2016, \apj,
  833, 102, \dodoi{10.3847/1538-4357/833/1/102}

\bibitem[{Billingsley(1995)}]{Billingsley1995}
Billingsley, P. 1995, {Probability and Measure}, 3rd edn. (New York: John Wiley
  {\&} Sons), 357--362

\bibitem[{{Bretteil} \& {Weber}(2005)}]{Bretteil2005}
{Bretteil}, S., \& {Weber}, R. 2005, Radio Science, 40, RS5S15,
  \dodoi{10.1029/2004RS003124}

\bibitem[{Burd {et~al.}(2018)Burd, Mannheim, MÃ€rz, Ringholz, Kappes, \&
  Kadler}]{Burd2018}
Burd, P.~R., Mannheim, K., MÃ€rz, T., {et~al.} 2018, Astronomische
  Nachrichten, 339, 358, \dodoi{10.1002/asna.201813505}

\bibitem[{{DeBoer} {et~al.}(2017){DeBoer}, {Parsons}, {Aguirre}, {Alexander},
  {Ali}, {Beardsley}, {Bernardi}, {Bowman}, {Bradley}, {Carilli}, {Cheng}, {de
  Lera Acedo}, {Dillon}, {Ewall-Wice}, {Fadana}, {Fagnoni}, {Fritz},
  {Furlanetto}, {Glendenning}, {Greig}, {Grobbelaar}, {Hazelton}, {Hewitt},
  {Hickish}, {Jacobs}, {Julius}, {Kariseb}, {Kohn}, {Lekalake}, {Liu}, {Loots},
  {MacMahon}, {Malan}, {Malgas}, {Maree}, {Martinot}, {Mathison}, {Matsetela},
  {Mesinger}, {Morales}, {Neben}, {Patra}, {Pieterse}, {Pober}, {Razavi-Ghods},
  {Ringuette}, {Robnett}, {Rosie}, {Sell}, {Smith}, {Syce}, {Tegmark},
  {Thyagarajan}, {Williams}, \& {Zheng}}]{Deboer2017}
{DeBoer}, D.~R., {Parsons}, A.~R., {Aguirre}, J.~E., {et~al.} 2017, \pasp, 129,
  045001, \dodoi{10.1088/1538-3873/129/974/045001}

\bibitem[{Furlanetto {et~al.}(2006)Furlanetto, Oh, \& Briggs}]{Furlanetto2006}
Furlanetto, S.~R., Oh, S.~P., \& Briggs, F.~H. 2006, Physics Reports, 433, 181
  , \dodoi{https://doi.org/10.1016/j.physrep.2006.08.002}

\bibitem[{Hellbourg {et~al.}(2012)Hellbourg, Weber, Capdessus, \&
  Boonstra}]{HELLBOURG2012}
Hellbourg, G., Weber, R., Capdessus, C., \& Boonstra, A.-J. 2012, Comptes
  Rendus Physique, 13, 71 , \dodoi{https://doi.org/10.1016/j.crhy.2011.10.010}

\bibitem[{{Hurley-Walker} {et~al.}(2017){Hurley-Walker}, {Callingham},
  {Hancock}, {Franzen}, {Hindson}, {Kapi{\'n}ska}, {Morgan}, {Offringa},
  {Wayth}, {Wu}, {Zheng}, {Murphy}, {Bell}, {Dwarakanath}, {For}, {Gaensler},
  {Johnston-Hollitt}, {Lenc}, {Procopio}, {Staveley-Smith}, {Ekers}, {Bowman},
  {Briggs}, {Cappallo}, {Deshpande}, {Greenhill}, {Hazelton}, {Kaplan},
  {Lonsdale}, {McWhirter}, {Mitchell}, {Morales}, {Morgan}, {Oberoi}, {Ord},
  {Prabu}, {Shankar}, {Srivani}, {Subrahmanyan}, {Tingay}, {Webster},
  {Williams}, \& {Williams}}]{Hurley-Walker2017}
{Hurley-Walker}, N., {Callingham}, J.~R., {Hancock}, P.~J., {et~al.} 2017,
  \mnras, 464, 1146, \dodoi{10.1093/mnras/stw2337}

\bibitem[{Kerrigan {et~al.}(2019)Kerrigan, Plante, Kohn, Pober, Aguirre,
  Abdurashidova, Alexander, Ali, Balfour, Beardsley, Bernardi, Bowman, Bradley,
  Burba, Carilli, Cheng, DeBoer, Dexter, de~Lera~Acedo, Dillon, Estrada,
  Ewall-Wice, Fagnoni, Fritz, Furlanetto, Glendenning, Greig, Grobbelaar,
  Gorthi, Halday, Hazelton, Hickish, Jacobs, Julius, Kern, Kittiwisit,
  Kolopanis, Lanman, Lekalake, Liu, MacMahon, Malan, Malgas, Maree, Martinot,
  Matsetela, Mesinger, Molewa, Morales, Mosiane, Neben, Parsons, Patra,
  Pieterse, Razavi-Ghods, Ringuette, Robnett, Rosie, Sims, Smith, Syce,
  Thyagarajan, Williams, \& Zheng}]{Kerrigan2019}
Kerrigan, J., Plante, P.~L., Kohn, S., {et~al.} 2019, Monthly Notices of the
  Royal Astronomical Society, \dodoi{10.1093/mnras/stz1865}

\bibitem[{{Loi} {et~al.}(2016){Loi}, {Cairns}, {Murphy}, {Erickson}, {Bell},
  {Rowlinson}, {Arora}, {Morgan}, {Ekers}, {Hurley-Walker}, \&
  {Kaplan}}]{Loi2016}
{Loi}, S.~T., {Cairns}, I.~H., {Murphy}, T., {et~al.} 2016, Journal of
  Geophysical Research (Space Physics), 121, 1569, \dodoi{10.1002/2015JA022052}

\bibitem[{{McKinley} {et~al.}(2013){McKinley}, {Briggs}, {Kaplan}, {Greenhill},
  {Bernardi}, {Bowman}, {de Oliveira-Costa}, {Tingay}, {Gaensler}, {Oberoi},
  {Johnston-Hollitt}, {Arcus}, {Barnes}, {Bunton}, {Cappallo}, {Corey},
  {Deshpande}, {deSouza}, {Emrich}, {Goeke}, {Hazelton}, {Herne}, {Hewitt},
  {Kasper}, {Kincaid}, {Koenig}, {Kratzenberg}, {Lonsdale}, {Lynch},
  {McWhirter}, {Mitchell}, {Morales}, {Morgan}, {Ord}, {Pathikulangara},
  {Prabu}, {Remillard}, {Rogers}, {Roshi}, {Salah}, {Sault}, {Udaya Shankar},
  {Srivani}, {Stevens}, {Subrahmanyan}, {Wayth}, {Waterson}, {Webster},
  {Whitney}, {Williams}, {Williams}, \& {Wyithe}}]{McKinley2013}
{McKinley}, B., {Briggs}, F., {Kaplan}, D.~L., {et~al.} 2013, \aj, 145, 23,
  \dodoi{10.1088/0004-6256/145/1/23}

\bibitem[{{McKinley} {et~al.}(2018){McKinley}, {Bernardi}, {Trott}, {Line},
  {Wayth}, {Offringa}, {Pindor}, {Jordan}, {Sokolowski}, {Tingay}, {Lenc},
  {Hurley-Walker}, {Bowman}, {Briggs}, \& {Webster}}]{McKinley2018}
{McKinley}, B., {Bernardi}, G., {Trott}, C.~M., {et~al.} 2018, \mnras, 481,
  5034, \dodoi{10.1093/mnras/sty2437}

\bibitem[{{Morales} \& {Wyithe}(2010)}]{Morales2010}
{Morales}, M.~F., \& {Wyithe}, J. S.~B. 2010, \araa, 48, 127,
  \dodoi{10.1146/annurev-astro-081309-130936}

\bibitem[{{Offringa} {et~al.}(2010){Offringa}, {de Bruyn}, {Biehl}, {Zaroubi},
  {Bernardi}, \& {Pandey}}]{Offringa2010}
{Offringa}, A.~R., {de Bruyn}, A.~G., {Biehl}, M., {et~al.} 2010, \mnras, 405,
  155, \dodoi{10.1111/j.1365-2966.2010.16471.x}

\bibitem[{Offringa {et~al.}(2012a)Offringa, de~Bruyn, \&
  Zaroubi}]{Offringa2012}
Offringa, A.~R., de~Bruyn, A.~G., \& Zaroubi, S. 2012a, Monthly Notices of the
  Royal Astronomical Society, 422, 563,
  \dodoi{10.1111/j.1365-2966.2012.20633.x}

\bibitem[{{Offringa} {et~al.}(2012b){Offringa}, {van de Gronde, J. J.}, \&
  {Roerdink, J. B. T. M.}}]{O2012}
{Offringa}, A.~R., {van de Gronde, J. J.}, \& {Roerdink, J. B. T. M.} 2012b,
  A\&A, 539, A95, \dodoi{10.1051/0004-6361/201118497}

\bibitem[{{Offringa} {et~al.}(2013){Offringa}, {de Bruyn}, {Zaroubi},
  {Koopmans}, {Wijnholds}, {Abdalla}, {Brouw}, {Ciardi}, {Iliev}, {Harker},
  {Mellema}, {Bernardi}, {Zarka}, {Ghosh}, {Alexov}, {Anderson}, {Asgekar},
  {Avruch}, {Beck}, {Bell}, {Bell}, {Bentum}, {Best}, {B{\^\i}rzan},
  {Breitling}, {Broderick}, {Br{\"u}ggen}, {Butcher}, {de Gasperin}, {de Geus},
  {de Vos}, {Duscha}, {Eisl{\"o}ffel}, {Fallows}, {Ferrari}, {Frieswijk},
  {Garrett}, {Grie{\ss}meier}, {Hassall}, {Horneffer}, {Iacobelli}, {Juette},
  {Karastergiou}, {Klijn}, {Kondratiev}, {Kuniyoshi}, {Kuper}, {van Leeuwen},
  {Loose}, {Maat}, {Macario}, {Mann}, {McKean}, {Meulman}, {Norden}, {Orru},
  {Paas}, {Pand ey-Pommier}, {Pizzo}, {Polatidis}, {Rafferty}, {Reich}, {van
  Nieuwpoort}, {R{\"o}ttgering}, {Scaife}, {Sluman}, {Smirnov}, {Sobey},
  {Tagger}, {Tang}, {Tasse}, {Veen}, {Toribio}, {Vermeulen}, {Vocks}, {van
  Weeren}, {Wise}, \& {Wucknitz}}]{Offringa2013}
{Offringa}, A.~R., {de Bruyn}, A.~G., {Zaroubi}, S., {et~al.} 2013, \mnras,
  435, 584, \dodoi{10.1093/mnras/stt1337}

\bibitem[{{Offringa} {et~al.}(2015){Offringa}, {Wayth}, {Hurley-Walker},
  {Kaplan}, {Barry}, {Beardsley}, {Bell}, {Bernardi}, {Bowman}, {Briggs},
  {Callingham}, {Cappallo}, {Carroll}, {Deshpand e}, {Dillon}, {Dwarakanath},
  {Ewall-Wice}, {Feng}, {For}, {Gaensler}, {Greenhill}, {Hancock}, {Hazelton},
  {Hewitt}, {Hindson}, {Jacobs}, {Johnston-Hollitt}, {Kapi{\'n}ska}, {Kim},
  {Kittiwisit}, {Lenc}, {Line}, {Loeb}, {Lonsdale}, {McKinley}, {McWhirter},
  {Mitchell}, {Morales}, {Morgan}, {Morgan}, {Neben}, {Oberoi}, {Ord}, {Paul},
  {Pindor}, {Pober}, {Prabu}, {Procopio}, {Riding}, {Udaya Shankar}, {Sethi},
  {Srivani}, {Staveley-Smith}, {Subrahmanyan}, {Sullivan}, {Tegmark},
  {Thyagarajan}, {Tingay}, {Trott}, {Webster}, {Williams}, {Williams}, {Wu},
  {Wyithe}, \& {Zheng}}]{Offringa2015}
{Offringa}, A.~R., {Wayth}, R.~B., {Hurley-Walker}, N., {et~al.} 2015, \pasa,
  32, e008, \dodoi{10.1017/pasa.2015.7}

\bibitem[{Peck \& Fenech(2013)}]{PECK2013}
Peck, L.~W., \& Fenech, D.~M. 2013, Astronomy and Computing, 2, 54 ,
  \dodoi{https://doi.org/10.1016/j.ascom.2013.09.001}

\bibitem[{{Sekhar} \& {Athreya}(2018)}]{Sekhar2018}
{Sekhar}, S., \& {Athreya}, R. 2018, \aj, 156, 9,
  \dodoi{10.3847/1538-3881/aac16e}

\bibitem[{{Sokolowski} {et~al.}(2016){Sokolowski}, {Wayth}, \&
  {Lewis}}]{Sokolowski2016}
{Sokolowski}, M., {Wayth}, R.~B., \& {Lewis}, M. 2016, arXiv e-prints,
  arXiv:1610.04696.
\newblock \doarXiv{1610.04696}

\bibitem[{{Sullivan} {et~al.}(2012){Sullivan}, {Morales}, {Hazelton}, {Arcus},
  {Barnes}, {Bernardi}, {Briggs}, {Bowman}, {Bunton}, {Cappallo}, {Corey},
  {Deshpande}, {deSouza}, {Emrich}, {Gaensler}, {Goeke}, {Greenhill}, {Herne},
  {Hewitt}, {Johnston-Hollitt}, {Kaplan}, {Kasper}, {Kincaid}, {Koenig},
  {Kratzenberg}, {Lonsdale}, {Lynch}, {McWhirter}, {Mitchell}, {Morgan},
  {Oberoi}, {Ord}, {Pathikulangara}, {Prabu}, {Remillard}, {Rogers}, {Roshi},
  {Salah}, {Sault}, {Udaya Shankar}, {Srivani}, {Stevens}, {Subrahmanyan},
  {Tingay}, {Wayth}, {Waterson}, {Webster}, {Whitney}, {Williams}, {Williams},
  \& {Wyithe}}]{Sullivan2012}
{Sullivan}, I.~S., {Morales}, M.~F., {Hazelton}, B.~J., {et~al.} 2012, \apj,
  759, 17, \dodoi{10.1088/0004-637X/759/1/17}

\bibitem[{{Taylor} {et~al.}(2012){Taylor}, {Ellingson}, {Kassim}, {Craig},
  {Dowell}, {Wolfe}, {Hartman}, {Bernardi}, {Clarke}, {Cohen}, {Dalal},
  {Erickson}, {Hicks}, {Greenhill}, {Jacoby}, {Lane}, {Lazio}, {Mitchell},
  {Navarro}, {Ord}, {Pihlstr{\"o}m}, {Polisensky}, {Ray}, {Rickard},
  {Schinzel}, {Schmitt}, {Sigman}, {Soriano}, {Stewart}, {Stovall}, {Tremblay},
  {Wang}, {Weiler}, {White}, \& {Wood}}]{Taylor2012}
{Taylor}, G.~B., {Ellingson}, S.~W., {Kassim}, N.~E., {et~al.} 2012, Journal of
  Astronomical Instrumentation, 1, 1250004, \dodoi{10.1142/S2251171712500043}

\bibitem[{{Tingay} {et~al.}(2013){Tingay}, {Goeke}, {Bowman}, {Emrich}, {Ord},
  {Mitchell}, {Morales}, {Booler}, {Crosse}, {Wayth}, {Lonsdale}, {Tremblay},
  {Pallot}, {Colegate}, {Wicenec}, {Kudryavtseva}, {Arcus}, {Barnes},
  {Bernardi}, {Briggs}, {Burns}, {Bunton}, {Cappallo}, {Corey}, {Deshpande},
  {Desouza}, {Gaensler}, {Greenhill}, {Hall}, {Hazelton}, {Herne}, {Hewitt},
  {Johnston-Hollitt}, {Kaplan}, {Kasper}, {Kincaid}, {Koenig}, {Kratzenberg},
  {Lynch}, {Mckinley}, {Mcwhirter}, {Morgan}, {Oberoi}, {Pathikulangara},
  {Prabu}, {Remillard}, {Rogers}, {Roshi}, {Salah}, {Sault}, {Udaya-Shankar},
  {Schlagenhaufer}, {Srivani}, {Stevens}, {Subrahmanyan}, {Waterson},
  {Webster}, {Whitney}, {Williams}, {Williams}, \& {Wyithe}}]{Tingay2012}
{Tingay}, S.~J., {Goeke}, R., {Bowman}, J.~D., {et~al.} 2013, \pasa, 30, e007,
  \dodoi{10.1017/pasa.2012.007}

\bibitem[{{van Haarlem} {et~al.}(2013){van Haarlem}, {Wise}, {Gunst}, {Heald},
  {McKean}, {Hessels}, {de Bruyn}, {Nijboer}, {Swinbank}, {Fallows},
  {Brentjens}, {Nelles}, {Beck}, {Falcke}, {Fender}, {H{\"o}randel},
  {Koopmans}, {Mann}, {Miley}, {R{\"o}ttgering}, {Stappers}, {Wijers},
  {Zaroubi}, {van den Akker}, {Alexov}, {Anderson}, {Anderson}, {van Ardenne},
  {Arts}, {Asgekar}, {Avruch}, {Batejat}, {B{\"a}hren}, {Bell}, {Bell}, {van
  Bemmel}, {Bennema}, {Bentum}, {Bernardi}, {Best}, {B{\^\i}rzan}, {Bonafede},
  {Boonstra}, {Braun}, {Bregman}, {Breitling}, {van de Brink}, {Broderick},
  {Broekema}, {Brouw}, {Br{\"u}ggen}, {Butcher}, {van Cappellen}, {Ciardi},
  {Coenen}, {Conway}, {Coolen}, {Corstanje}, {Damstra}, {Davies}, {Deller},
  {Dettmar}, {van Diepen}, {Dijkstra}, {Donker}, {Doorduin}, {Dromer}, {Drost},
  {van Duin}, {Eisl{\"o}ffel}, {van Enst}, {Ferrari}, {Frieswijk}, {Gankema},
  {Garrett}, {de Gasperin}, {Gerbers}, {de Geus}, {Grie{\ss}meier}, {Grit},
  {Gruppen}, {Hamaker}, {Hassall}, {Hoeft}, {Holties}, {Horneffer}, {van der
  Horst}, {van Houwelingen}, {Huijgen}, {Iacobelli}, {Intema}, {Jackson},
  {Jelic}, {de Jong}, {Juette}, {Kant}, {Karastergiou}, {Koers}, {Kollen},
  {Kondratiev}, {Kooistra}, {Koopman}, {Koster}, {Kuniyoshi}, {Kramer},
  {Kuper}, {Lambropoulos}, {Law}, {van Leeuwen}, {Lemaitre}, {Loose}, {Maat},
  {Macario}, {Markoff}, {Masters}, {McFadden}, {McKay-Bukowski}, {Meijering},
  {Meulman}, {Mevius}, {Middelberg}, {Millenaar}, {Miller-Jones}, {Mohan},
  {Mol}, {Morawietz}, {Morganti}, {Mulcahy}, {Mulder}, {Munk}, {Nieuwenhuis},
  {van Nieuwpoort}, {Noordam}, {Norden}, {Noutsos}, {Offringa}, {Olofsson},
  {Omar}, {Orr{\'u}}, {Overeem}, {Paas}, {Pand ey-Pommier}, {Pandey}, {Pizzo},
  {Polatidis}, {Rafferty}, {Rawlings}, {Reich}, {de Reijer}, {Reitsma},
  {Renting}, {Riemers}, {Rol}, {Romein}, {Roosjen}, {Ruiter}, {Scaife}, {van
  der Schaaf}, {Scheers}, {Schellart}, {Schoenmakers}, {Schoonderbeek},
  {Serylak}, {Shulevski}, {Sluman}, {Smirnov}, {Sobey}, {Spreeuw}, {Steinmetz},
  {Sterks}, {Stiepel}, {Stuurwold}, {Tagger}, {Tang}, {Tasse}, {Thomas},
  {Thoudam}, {Toribio}, {van der Tol}, {Usov}, {van Veelen}, {van der Veen},
  {ter Veen}, {Verbiest}, {Vermeulen}, {Vermaas}, {Vocks}, {Vogt}, {de Vos},
  {van der Wal}, {van Weeren}, {Weggemans}, {Weltevrede}, {White}, {Wijnholds},
  {Wilhelmsson}, {Wucknitz}, {Yatawatta}, {Zarka}, {Zensus}, \& {van
  Zwieten}}]{vanHaarlem2013}
{van Haarlem}, M.~P., {Wise}, M.~W., {Gunst}, A.~W., {et~al.} 2013, \aap, 556,
  A2, \dodoi{10.1051/0004-6361/201220873}

\bibitem[{{Wayth} {et~al.}(2018){Wayth}, {Tingay}, {Trott}, {Emrich},
  {Johnston-Hollitt}, {McKinley}, {Gaensler}, {Beardsley}, {Booler}, {Crosse},
  {Franzen}, {Horsley}, {Kaplan}, {Kenney}, {Morales}, {Pallot}, {Sleap},
  {Steele}, {Walker}, {Williams}, {Wu}, {Cairns}, {Filipovic}, {Johnston},
  {Murphy}, {Quinn}, {Staveley-Smith}, {Webster}, \& {Wyithe}}]{Wayth2018}
{Wayth}, R.~B., {Tingay}, S.~J., {Trott}, C.~M., {et~al.} 2018, \pasa, 35, 33,
  \dodoi{10.1017/pasa.2018.37}

\bibitem[{{Zhang} {et~al.}(2018){Zhang}, {Hancock}, {Devillepoix}, {Wayth},
  {Beardsley}, {Crosse}, {Emrich}, {Franzen}, {Gaensler}, {Horsley},
  {Johnston-Hollitt}, {Kaplan}, {Kenney}, {Morales}, {Pallot}, {Steele},
  {Tingay}, {Trott}, {Walker}, {Williams}, {Wu}, {Ji}, \& {Ma}}]{Zhang2018}
{Zhang}, X., {Hancock}, P., {Devillepoix}, H.~A.~R., {et~al.} 2018, \mnras,
  477, 5167, \dodoi{10.1093/mnras/sty930}

\bibitem[{Zwillinger \& Kokoska(2000)}]{Zwillinger2000}
Zwillinger, D., \& Kokoska, S. 2000, {Standard Probability and Statistics
  Tables and Formulae} (Boca Raton: CRC Press LLC), 55

\end{thebibliography}

\end{document}